\PassOptionsToPackage{numbers,sort&compress}{natbib}  
\documentclass[preprint,12pt]{elsarticle}
\usepackage[margin=2.5cm]{geometry}
\usepackage{enumitem}
\newlist{inlinelist}{enumerate*}{1}
\setlist*[inlinelist,1]{%
  label=(\roman*),
}
\usepackage{soul}
\usepackage{hhline}
\usepackage{subfig}
\usepackage{multirow}
\usepackage{xcolor}
\usepackage{graphicx}
\usepackage{upgreek}
\usepackage{amssymb}
\usepackage{amsfonts,amsthm,bm,amsmath} 
\usepackage{appendix}
\usepackage{times}
\usepackage{caption}
\usepackage{subfig}
\newcommand{\psubref}[1]{\protect\subref{#1}}
\usepackage{multirow}
\usepackage{xcolor}
\usepackage[linesnumbered,ruled,vlined]{algorithm2e}
\usepackage{tablefootnote}

\SetCommentSty{mycommfont}

\SetKwInput{KwInput}{Input}                
\SetKwInput{KwOutput}{Output}              

\usepackage[colorlinks]{hyperref} 
\hypersetup{ 
    colorlinks=true,       
    linkcolor=red,          
    citecolor=blue,        
    filecolor=magenta,      
    urlcolor=cyan           
}

\newcommand{\fref}[1]{Fig.~\ref{#1}}

\newcommand{\eref}[1]{Eq.~(\ref{#1})}

\newcommand{\sref}[1]{Section~\ref{#1}}
\newcommand{\tref}[1]{Table~\ref{#1}}
\setcitestyle{square}

\journal{Computer Methods in Applied Mechanics and Engineering}
\begin{document}

\begin{frontmatter}

\title{Novel DeepONet architecture to predict stresses in elastoplastic structures with variable complex geometries and loads}
\author[]{Junyan He$^1$}
\author[]{Seid Koric$^{1,2}$}
\author[]{Shashank Kushwaha$^{1}$}
\author[]{Jaewan Park$^1$}
\author[]{Diab Abueidda$^{2,3}$}
\author[]{Iwona Jasiuk$^1$\corref{mycorrespondingauthor}}
\cortext[mycorrespondingauthor]{Corresponding author}
\ead{ijasiuk@illinois.edu}
\address{$^1$ Department of Mechanical Science and Engineering, University of Illinois at Urbana-Champaign, Champaign, IL, USA \\
$^2$ National Center for Supercomputing Applications, University of Illinois at Urbana-Champaign, Champaign, IL, USA \\
$^3$ Civil and Urban Engineering Department, New York University Abu Dhabi, Abu Dhabi, UAE\\
}

\begin{abstract}
A novel deep operator network (DeepONet) with a residual U-Net (ResUNet) as the trunk network is devised to predict full-field highly nonlinear elastic-plastic stress response for complex geometries obtained from topology optimization under variable loads. The proposed DeepONet uses a ResUNet in the trunk to encode complex input geometries, and a fully-connected branch network encodes the parametric loads. Additional information fusion is introduced via an element-wise multiplication of the encoded latent space to improve prediction accuracy further. The performance of the proposed DeepONet was compared to two baseline models, a standalone ResUNet and a DeepONet with fully connected networks as the branch and trunk. The results show that ResUNet and the proposed DeepONet share comparable accuracy; both can predict the stress field and accurately identify stress concentration points. However, the novel DeepONet is more memory efficient and allows greater flexibility with framework architecture modifications. The DeepONet with fully connected networks suffers from high prediction error due to its inability to effectively encode the complex, varying geometry. Once trained, all three networks can predict the full stress distribution orders of magnitude faster than finite element simulations. The proposed network can quickly guide preliminary optimization, designs, sensitivity analysis, uncertainty quantification, and many other nonlinear analyses that require extensive forward evaluations with variable geometries, loads, and other parameters. This work marks the first time a ResUNet is used as the trunk network in the DeepONet architecture and the first time that DeepONet solves problems with complex, varying input geometries under parametric loads and elasto-plastic material behavior.
\end{abstract}


\end{frontmatter}

\section{Introduction}
\label{sec:intro}
Machine learning and neural networks (NNs) have found wide applications in the field of computational mechanics. Challenging mechanics problems such as topology optimization (TO) \cite{zehnder2021ntopo,he2022deep,kollmann2020deep}, finite deformation \cite{egli2021surrogate,he2023use,abueidda2022deep}, path-dependent plastic deformation \cite{abueidda2021deep,he2023deep}, crack propagation \cite{schwarzer2019learning,yang2020prediction} and additive manufacturing \cite{perumal2023temporal,sadeghpour2022data} have been solved successfully by physics-informed or data-driven NNs. The universal approximation capability \cite{hornik1989multilayer} of NNs allows them to approximate intricate and complex solution fields that are often times computationally expensive to obtain via traditional numerical methods such as finite element (FE) and finite difference. Once a NN is trained, it can infer the full solution field at a speed several orders of magnitudes faster than traditional numerical methods \cite{he2023deep,koric2023deep,he2023exploring}. The ultra-fast inference speed could be beneficial in situations where many evaluations are needed with changing inputs. In engineering practice, it is common to seek the stress distribution of structures, whether designed by a designer or generated via topology optimization. These structures can be complex, having fine-scale structural details, especially since the advancement in additive manufacturing \cite{kellner2017epiphany} has enabled them to be manufactured. These engineering structures may also be under complex loading of variable magnitudes and directions applied to different locations of the structure. Furthermore, the material behavior of the structure can be complex, sometimes even exceeding the material's elastic limit. Stress prediction on variable geometries subject to changing loads is a particularly challenging problem, as different structural features induce vastly different local stress concentrations, and a small change in the design can significantly change the stress field. Therefore, in this paper, we developed and compared three data-driven NN models to predict the stress field on complex and varying input geometries subject to variable loads and an elastic-plastic material model. 

Past research efforts have formulated the stress prediction problem as image-to-image translation \cite{hoq2023data,nie2020stress,buehler2022end}, where the input image is the design and the output image is the full-field stress contour. As the designs are typically 2D or 3D that contain important spatial information, convolution neural networks (CNNs) \cite{lecun1989backpropagation} are widely used, such as the residual net \cite{he2016deep,hoq2023data}, the U-Net \cite{ronneberger2015u}, the residual U-Net (ResUNet) \cite{diakogiannis2020resunet,kollmann2020deep}, and the generative adversarial network \cite{goodfellow2020generative,buehler2022end,buehler2022prediction}. These network architectures typically consist of a convolutional encoder to extract and encode important spatial information in the input geometry and compress them to a compact latent space. A decoder then decodes the latent information back to its original dimension. CNNs are highly effective for images. However, they are limited to rectangular domains with a structured grid and can only generate predictions at the grid cell. Nie et al. \cite{nie2020stress} proposed two CNN-based NNs to predict stress distribution on different cantilever beam designs subject to variable loads with high accuracy. However, the FE mesh of the cantilever structures was extremely coarse (32$\times$24), so they lacked fine-scale design features that are common in TO. In addition, only 28 distinct designs were included in the training dataset with a linear elastic material model.

The stress prediction problem can also be formulated as a nonlinear operator learning task. The target operator (FE simulation) maps the input geometry and load information to the output stress field. For learning complex, nonlinear operators, the deep operator network (DeepONet) \cite{lu2021learning} is a recently proposed NN architecture that has shown tremendous success in approximating complex physics operators such as heat diffusion \cite{koric2023data,koric2022applications}, plastic deformation on a dogbone specimen \cite{koric2023deep}, multi-scale analysis \cite{yin2022interfacing}, crack propagation \cite{goswami2022physics}, Darcy flow over complex domains \cite{goswami2022deep} and engine combustion \cite{kumar2023real}. As proposed by Lu et al. \cite{lu2021learning}, a DeepONet consists of two different networks: the branch and the trunk. The branch network encodes one or multiple input source functions, while the trunk network encodes the geometry of the problem domain. Information from the two networks is then combined together via a dot product. Since the original DeepONet architecture, many modified versions exist, such as having a CNN in the branch \cite{oommen2022learning}, having multiple branch networks for multiple input source functions \cite{tan2022enhanced}, and having intermediate data fusion between the branch and trunk before the final dot product \cite{wang2022improved}. However, the application of DeepONet to solve engineering problems with variable geometries, variable loads, and nonlinear material models has been under-explored. Goswami et al. \cite{goswami2022deep} applied DeepONet and transfer learning to extend accurate predictions onto different geometries distinct from that in the training data set. For the first time in literature, Koric et al. applied DeepONet to predict the stress field on a rectangular domain with varying elastic-plastic material properties and parametric loads. They also applied DeepONet to predict the stress field in a complex dogbone geometry with time-dependent loads but did not explore the scenario with multiple complex geometries in the training dataset. Stress prediction on complex geometries, variable loads, and elastic-plastic material models is a complex task that has never been attempted before with DeepONet in the literature. Therefore, one of the objectives of this paper is to develop a novel DeepONet architecture capable of predicting the stress field under these complex input conditions.

This paper is organized as follows: \sref{sec:methods} provides detail on the data generation method and introduces three neural network architectures. \sref{sec:results} presents and discusses the performance of the three neural networks. \sref{sec:conc} summarizes the outcomes and highlights possible future works.

\section{Methods}
\label{sec:methods}
\subsection{Training data generation}
\label{sec:data_gen}
\begin{figure}[h!] 
    \centering
     \subfloat[]{
         \includegraphics[trim={0cm 0cm 0cm 0cm},clip,width=0.35\textwidth]{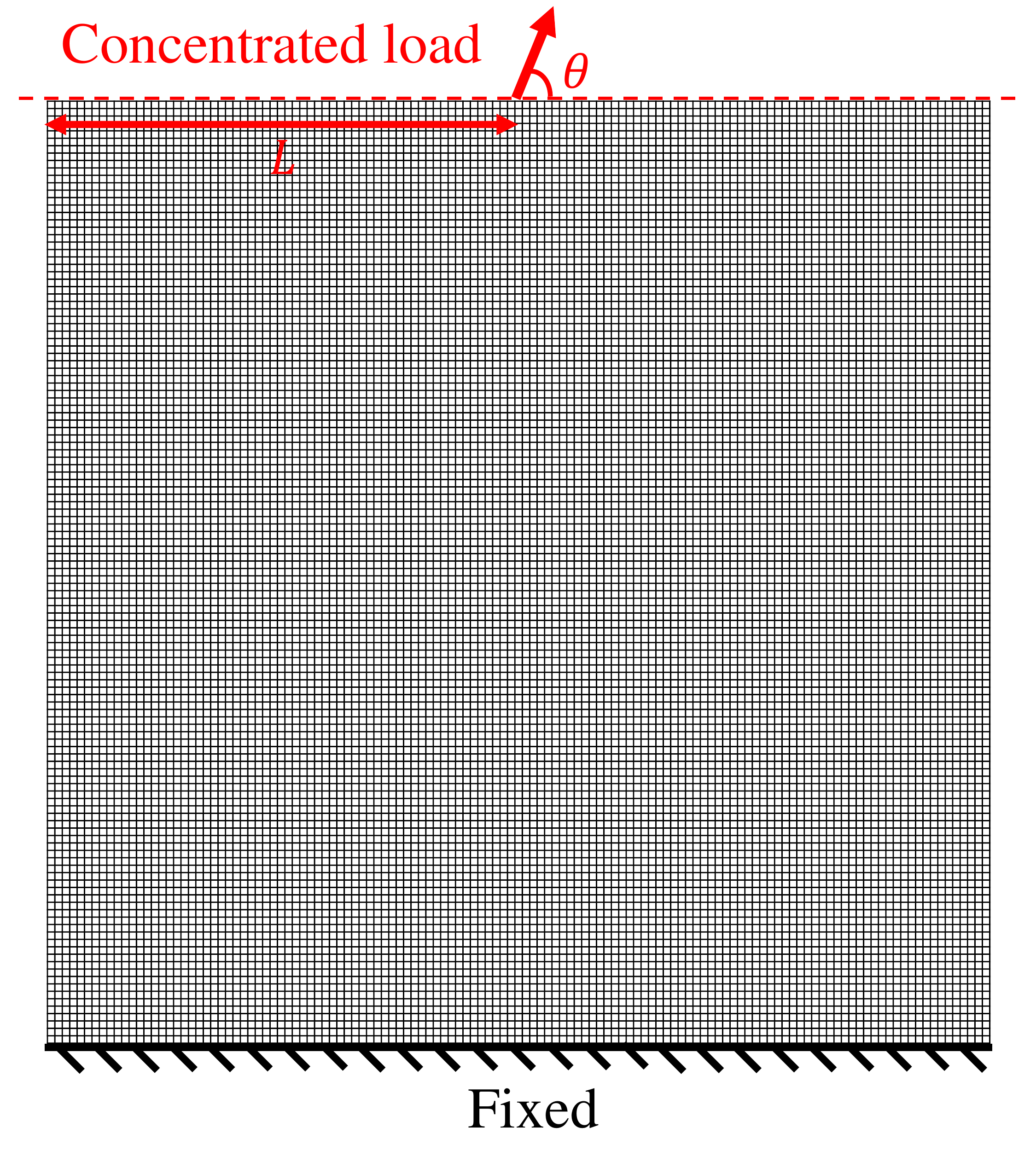}
         \label{mesh_and_geom}
     }
     \subfloat[]{
         \includegraphics[trim={0cm 0cm 0cm 0cm},clip,width=0.35\textwidth]{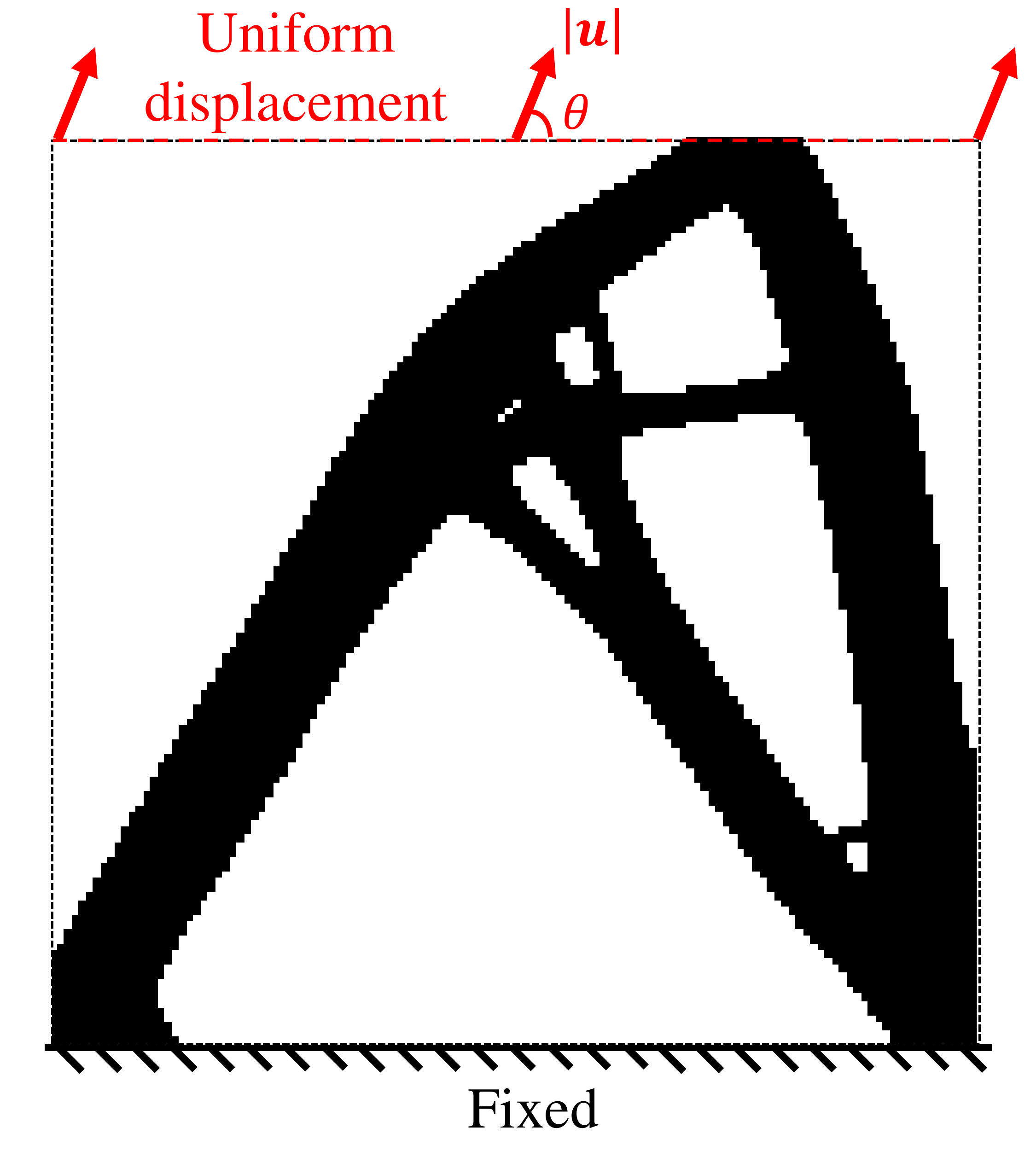}
         \label{to_geom}
     }
    \caption{FE simulation settings: \psubref{mesh_and_geom} Mesh, load, and boundary conditions used in elastic TO. \psubref{to_geom} Binary geometry used in elastic-plastic simulation and the load and boundary conditions.}
    \label{FE_setting}
\end{figure}

\begin{figure}[h!] 
    \centering
     \subfloat[ (0.31,0.08,3.75) ]{
         \includegraphics[trim={0cm 0cm 0cm 0cm},clip,width=0.19\textwidth]{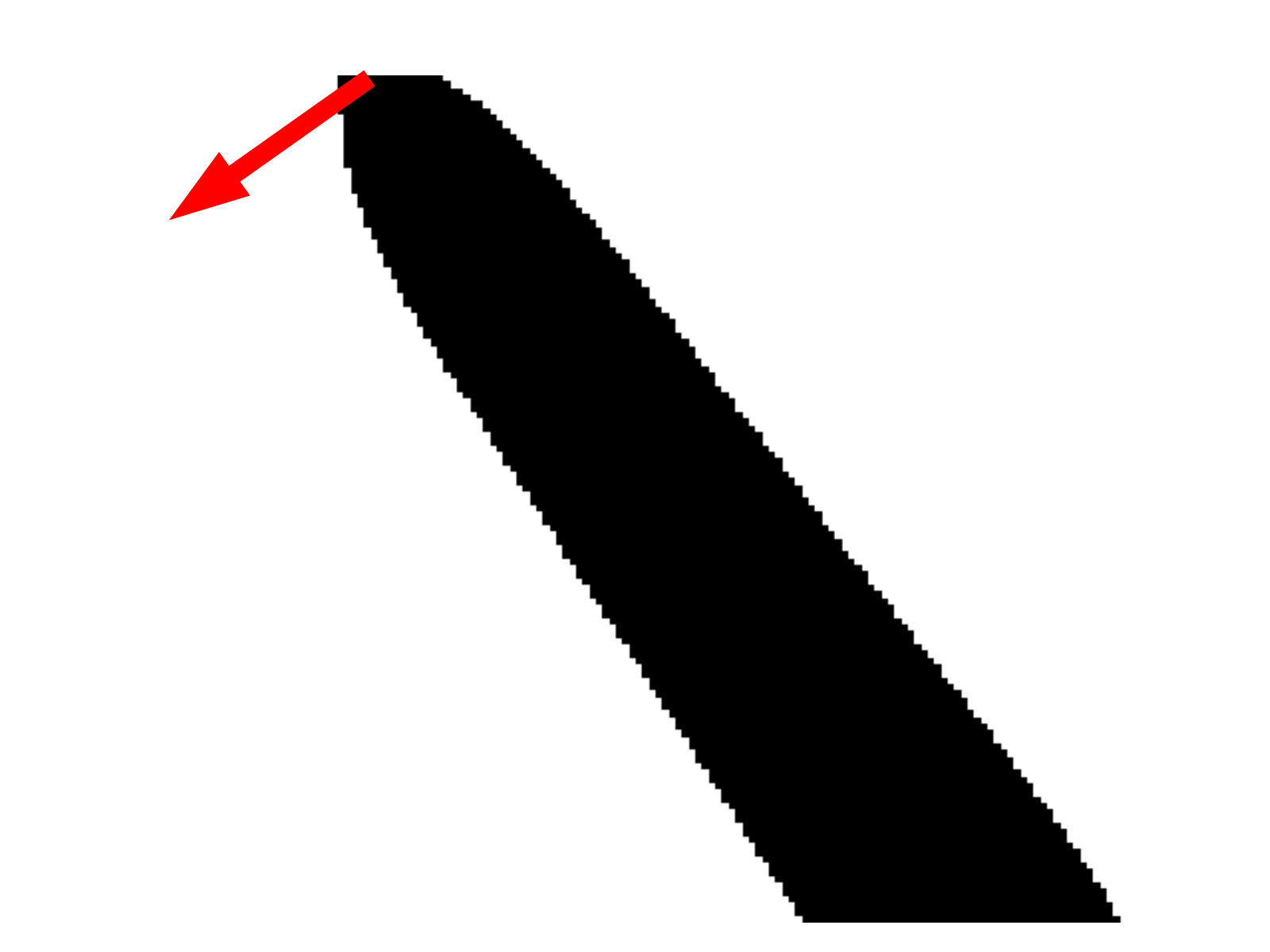}
         \label{d0}
     }
     \subfloat[ (0.23,0.71,1.27) ]{
         \includegraphics[trim={0cm 0cm 0cm 0cm},clip,width=0.19\textwidth]{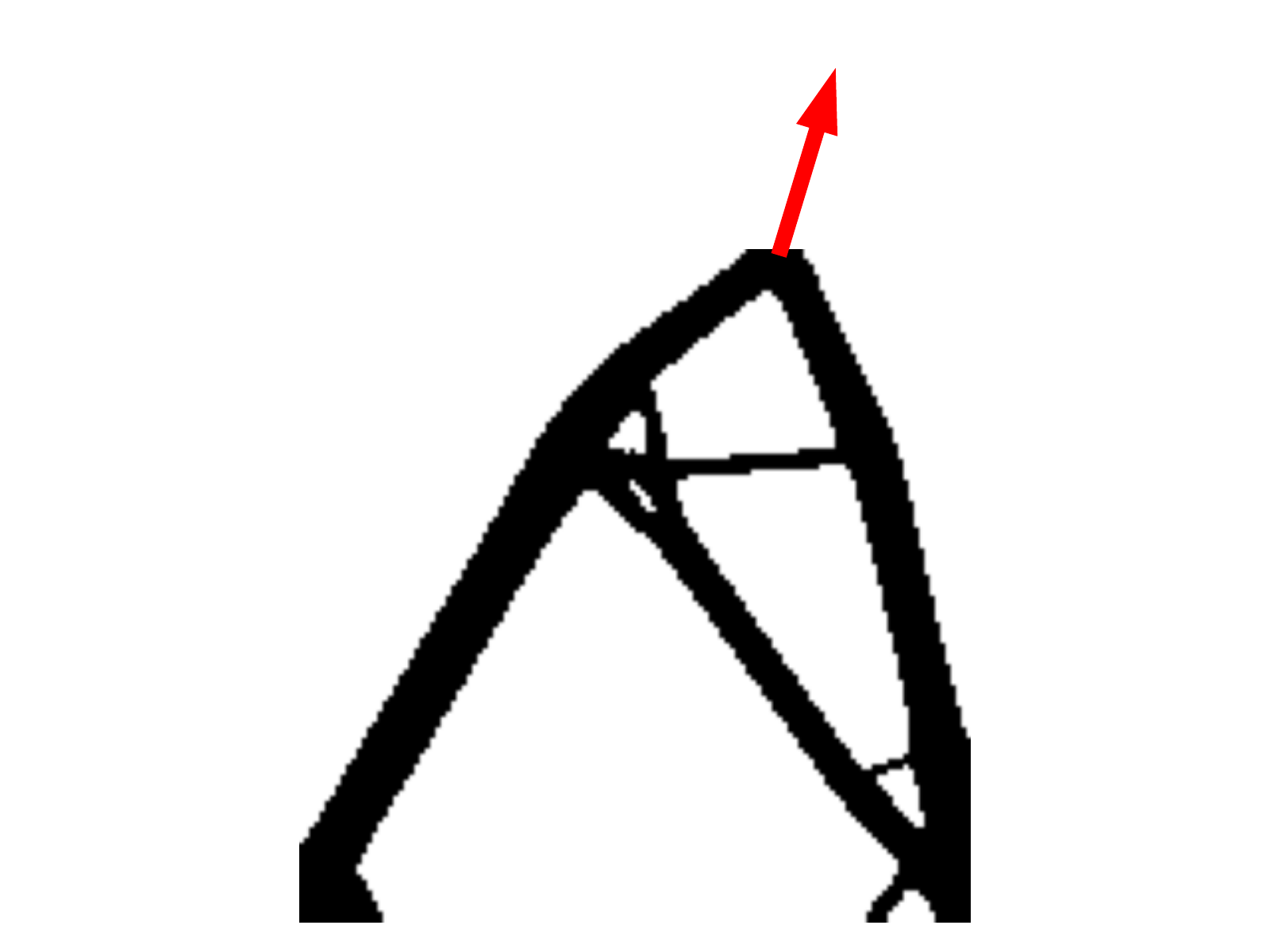}
         \label{d1}
     }
     \subfloat[ (0.23,0.85,0.59) ]{
         \includegraphics[trim={0cm 0cm 0cm 0cm},clip,width=0.19\textwidth]{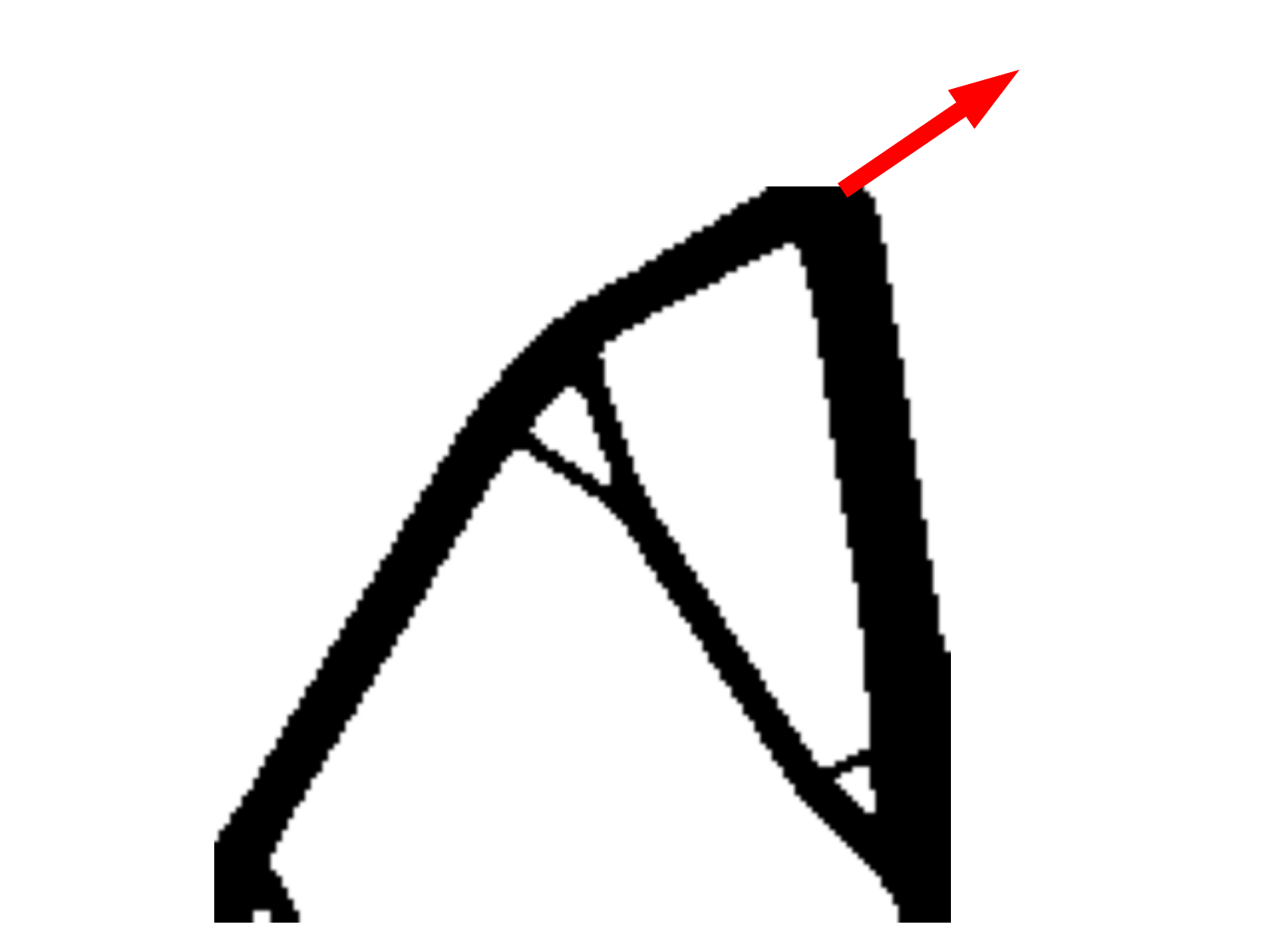}
         \label{d2}
     }
     \subfloat[ (0.53,0.15,5.11) ]{
         \includegraphics[trim={0cm 0cm 0cm 0cm},clip,width=0.19\textwidth]{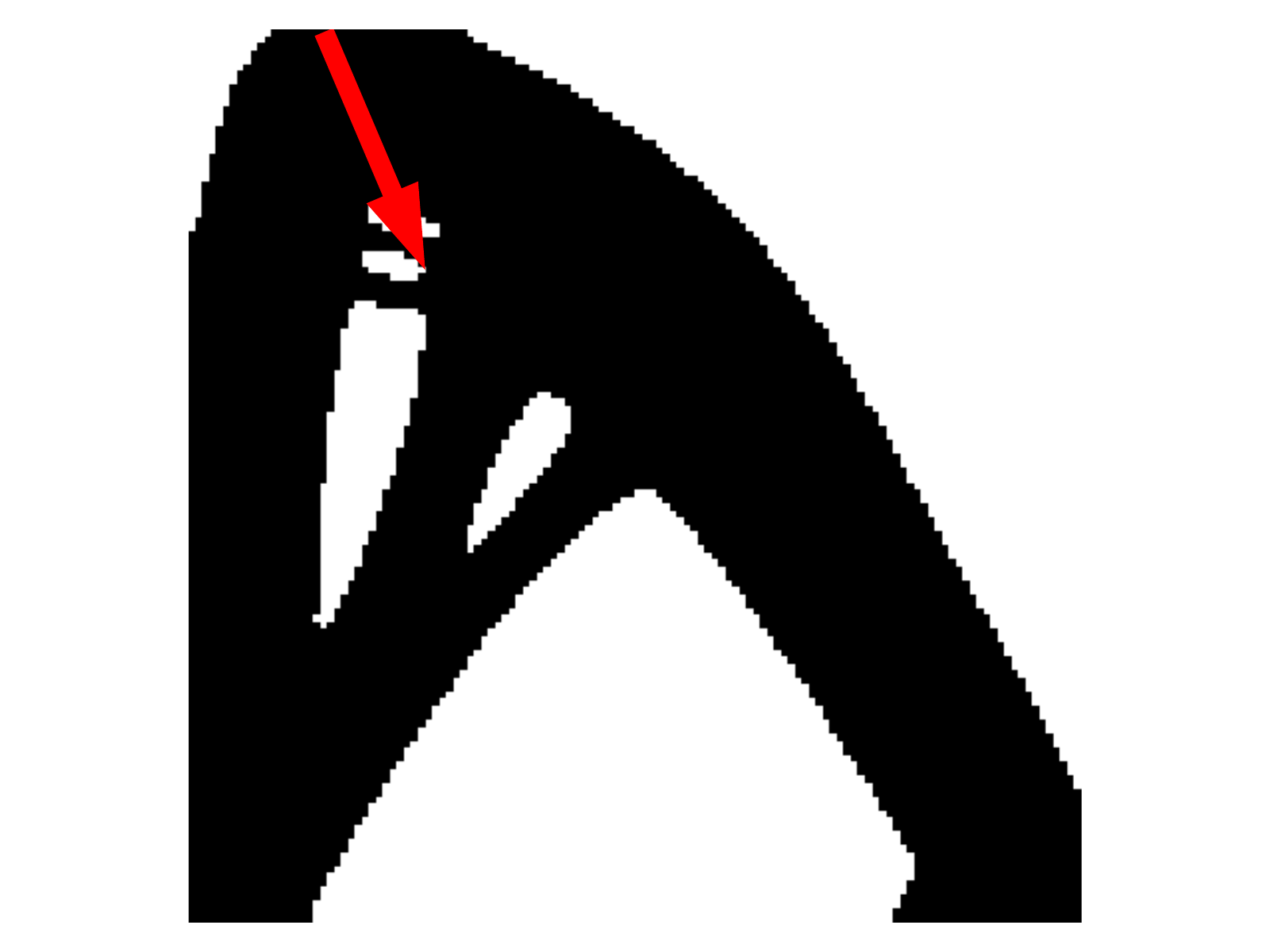}
         \label{d3}
     }
     \subfloat[ (0.32,0.07,4.06) ]{
         \includegraphics[trim={0cm 0cm 0cm 0cm},clip,width=0.19\textwidth]{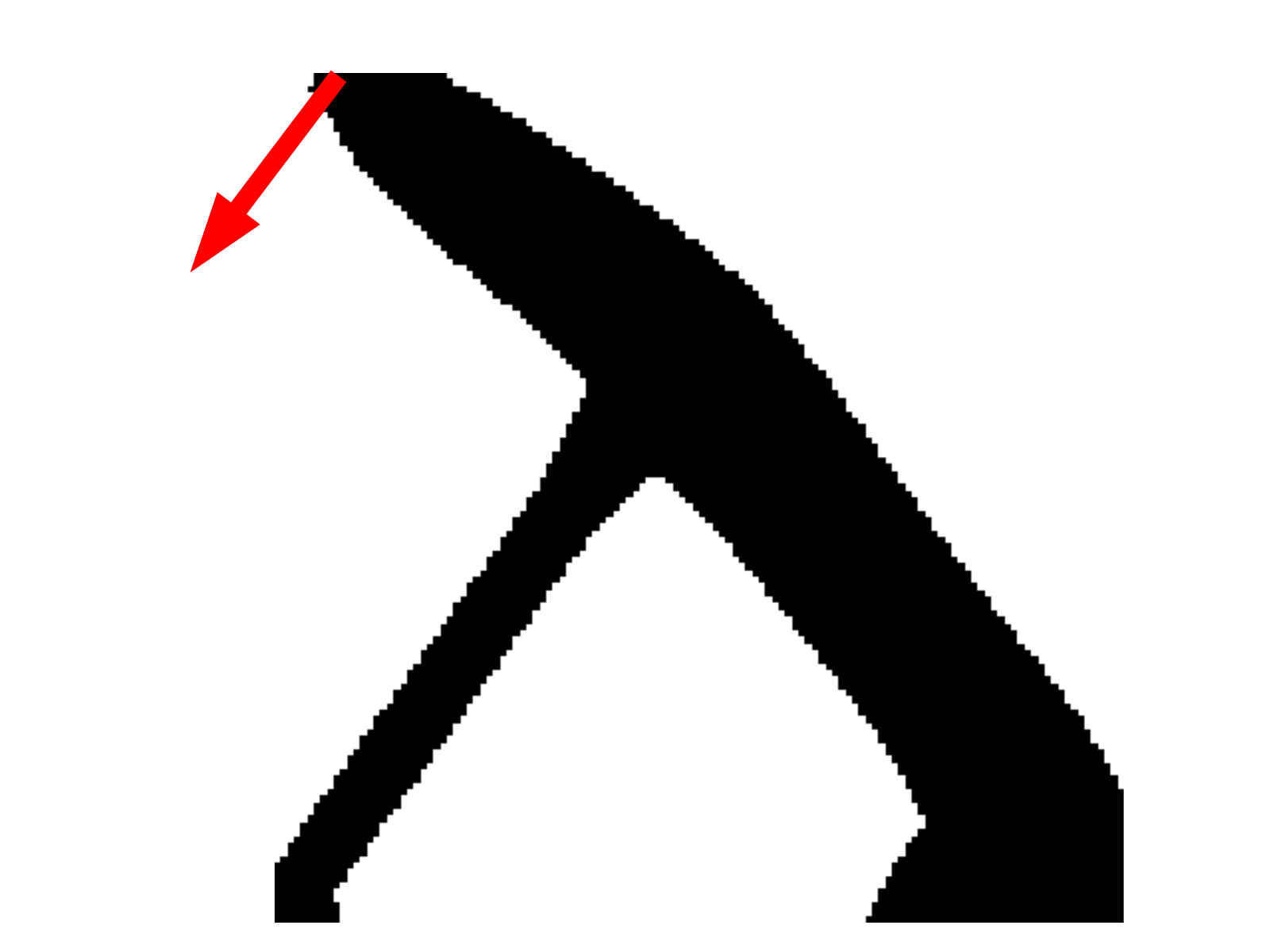}
         \label{d4}
     }
    \caption{Selected final designs generated by TO. TO cases are identified by a tuple of three parameters $(V_f,L,\theta)$. The direction of the applied load is shown in a red arrow.}
    \label{to_designs}
\end{figure}

A two-stage data generation method was employed to generate an extensive data set with complex, distinct geometries, variable loading, and elastic-plastic material behavior. In the first stage, a fully-elastic TO was conducted on a 128$\times$128 structured mesh with variable loads and design volume fractions. The mathematical statement of the TO problem is as follows:
\begin{equation}
\begin{aligned}
    \min_{ \bm{\rho} } \bm{F} \cdot \bm{u} \\
    {\rm{s.t.}} \; \bm{K} \bm{u} = \bm{F}, \\
    \sum \bm{\rho} = N \times V_f, \\
    0 \le \rho_e \le 1, \, \forall e = 1 \cdots N,
\end{aligned}
\end{equation}
where $\rho_e$, $N$, and $V_f$ denote the element density, total number of elements, and target volume fraction, respectively. The Python script used for the elastic TO was adapted from that by Aage \cite{top_opt}, which followed the work of Andreassen et al. \cite{andreassen2011efficient}. The mesh, load, and boundary conditions are shown in \fref{mesh_and_geom}. For each TO case, a concentrated load of unitary magnitude was applied on the top edge of the design domain. The load angle $\theta$ and normalized load location $L$ were randomly sampled from the ranges $[0,2\pi]$ and $[0,1]$, respectively. To further increase the diversity of the generated designs, the volume fraction of the final design $V_f$ was also randomly sampled in the range $[0.2,0.6]$. Each TO case can be uniquely identified by a tuple of three parameters $(V_f,L,\theta)$. Density filtering was used to prevent checker-boarding in the final designs with a filter radius of 2 element widths. A total of 30 TO iterations were conducted for each TO case, and a total of 3000 TO cases were completed to generate 3000 distinct designs. All final designs were binarized with an element density threshold of 0.5. To showcase the diverse final designs generated by the TO process, selected final designs are shown in \fref{to_designs}.

Once distinct input geometries have been generated from elastic TO, an additional simulation was conducted on each design using Abaqus \cite{Abaqus2021} featuring an elastic-plastic material model. In this work, the small-strain formulation of plasticity was used, where the total strain is decomposed additively into its elastic part $\bm{\epsilon}^e$ and plastic part $\bm{\epsilon}^p$:
\begin{equation}
    \bm{\epsilon} = \bm{\epsilon}^e + \bm{\epsilon}^p.
    \label{decomposition}
\end{equation}
For linear elastic and isotropic material under plane stress condition, the constitutive equation is:
\begin{equation}
    \begin{bmatrix}
    \sigma_{11} \\
    \sigma_{22} \\
    \sigma_{12}
    \end{bmatrix}
    =
    \begin{bmatrix}
    \frac{E}{1-\nu^2} & \frac{\nu E}{1-\nu^2} & 0 \\
    \frac{\nu E}{1-\nu^2} & \frac{E}{1-\nu^2} & 0 \\
    0 & 0 & \frac{E}{2(1+\nu)}
    \end{bmatrix}
    \begin{bmatrix}
    \epsilon_{11} \\
    \epsilon_{22} \\
    \epsilon_{12}
    \end{bmatrix},
    \label{stress}
\end{equation}
where $E$ and $\nu$ are the Young's modulus and Poisson's ratio. In this work, $J_2$ plasticity with an associative flow rule is used, which assumes incompressible plastic strains. After the initial yield point, a linear isotropic hardening behavior is assumed; see \eref{hardening}:
\begin{equation}
    \sigma_y( \Bar{\epsilon}_p ) = \sigma_{y0} + H \Bar{\epsilon}_p,
    \label{hardening}
\end{equation}
where $\sigma_y$, $\Bar{\epsilon}_p$, $\sigma_{y0}$, and $H$ denote the flow stress, equivalent plastic strain, initial yield stress, and the hardening modulus, respectively. The material properties of the elastic-plastic material response are presented in \tref{mat_props}. The radial return algorithm \citep{wilkins1964methods} is used to update the plastic state variables based on the hardening rule. The elements corresponding to an element density of 0 in the binarized design were deactivated (i.e., not included in the FE global stiffness matrix) during the simulations to accelerate the simulation speed, and the output stress value is set to 0 for all deactivated elements to maintain a constant output size of 128$\times$128. 

\begin{table}[h!]
    \caption{Material properties of the elastic-plastic material model}
    \small
    \centering
    \begin{tabular}{cccccccc}
     Property & \vline & $E$ [MPa] & $\nu$ [/] & $\sigma_{y0}$ [MPa] & H [MPa]\\
    \hline
    Value & \vline  & 2.09$\times 10^{5}$ & 0.3 & 235 & 836\\
    \end{tabular}
    \label{mat_props}
\end{table}

For simplicity and to adapt to the varying geometries, a uniform displacement with variable angle $\theta$ and magnitude $|\bm{u}|$ was applied on the entire top edge of the domain, and the bottom edge of the domain remains fixed, as shown in \fref{to_geom}. The load angle $\theta$ and displacement magnitude were randomly sampled from the range $[0,\pi]$ and $[2,8]$, respectively. The displacement magnitude range was chosen such that some plastic deformation will occur in all the final designs. For each of the 3000 designs, five randomly generated load cases were simulated, leading to a total of 15000 simulations. The von Mises stress $\Bar{\sigma}$ at each element was stored as the ground truth labels in the NN training, and is defined as:
\begin{equation}
    \Bar{\sigma} = \sqrt{ \sigma_{11}^2 + \sigma_{22}^2 + \sigma_{11}\sigma_{22} + 3\sigma_{12}^2 }.
\end{equation}

\subsection{Neural network models}
\label{sec:NNs}
This work explored three different NN architectures to predict the final von Mises stress contour given the input element density $\bm{\rho}$ and the load parameters $\theta$ and $|\bm{u}|$. All three NNs were implemented in the DeepXDE framework \cite{lu2021deepxde} with a TensorFlow backend \cite{tensorflow2015-whitepaper}. 

\subsubsection{ResUNet}
\label{sec:resunet}
\begin{figure}[h!] 
    \centering
         \includegraphics[width=0.3\textwidth]{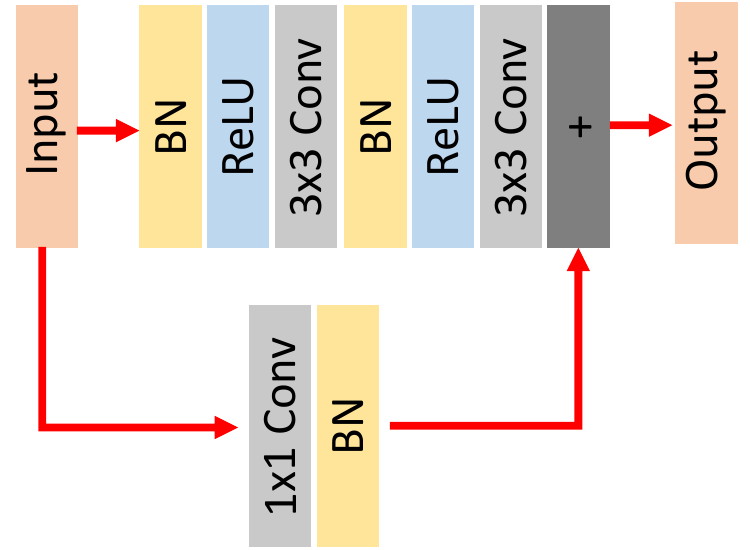}
    \caption{Schematic of a residual block used in the ResUNet.}
    \label{ResBlock}
\end{figure}

\begin{figure}[h!] 
    \centering
         \includegraphics[width=0.9\textwidth]{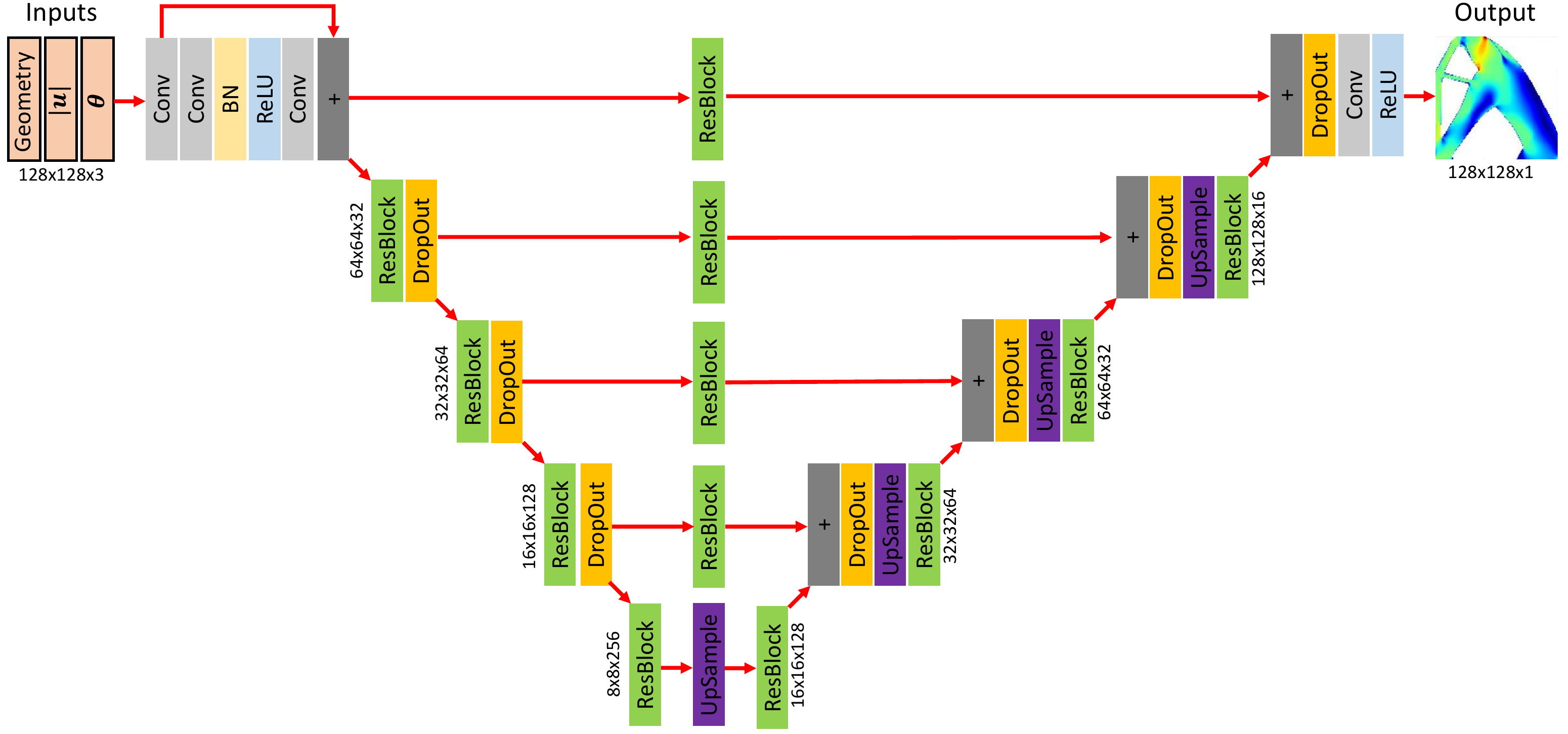}
    \caption{Schematic of the ResUNet used in this work.}
    \label{Resunet}
\end{figure}

The ResUNet \cite{diakogiannis2020resunet} is a CNN-based NN architecture commonly used for image semantic segmentation. It combines the residual learning via the residual block (see \fref{ResBlock}), and the U-Net architecture \cite{ronneberger2015u}. 
The Python implementation of ResUNet was modified from the original code in \cite{resunet} to replace the concatenation layers with "addition" layers during the encoding-decoding process to remove the significant overhead of the concatenation layers when running with TensorFlow's XLA acceleration. The proposed ResUNet architecture is shown in \fref{Resunet}, which takes 3 128$\times$128 images (Geometry, with $|\bm{u}|$ and $\theta$ as two constant images) as inputs and predicts the von Mises stress distribution. Batch normalization (BN) and ReLU activation layers were used in the residual blocks following the 3$\times$3 convolution layers. Drop-out layers with a drop-out rate of 0.02 were added during encoding and decoding to minimize overfitting. The proposed ResUNet had a total of 3569441 trainable parameters. The Adam \cite{kingma2014adam} optimizer was used with an initial learning rate of 5$\times 10^{-4}$ and an inverse time decay learning rate schedule with a decay rate of 1$\times 10^{-4}$. The model was trained for 150000 epochs with a batch size of 16 on 80\% of the total data, with 20\% reserved for testing. The scaled mean squared error (MSE) was used as the training loss function (output stresses normalized by a constant magnitude of 500 MPa), and the mean relative $L_2$ error (MRL2E) was used as the test metric:
\begin{equation}
    {\rm{MRL2E}} = \frac{1}{N_T} \sum_{i=1}^{N_T} \frac{ | \bm{\sigma}_{GT} - \bm{\sigma}_{Pred} * \bm{\rho} |_2 }{ |\bm{\sigma}_{GT}|_2 },
\end{equation}
where $N_T$, $\bm{\sigma}_{GT}$, $\bm{\sigma}_{Pred}$ and $\bm{\rho}$ denote the total number of test cases, the FE-simulated von Mises stress, the NN-predicted stress, and the binarized input element density, respectively. $*$ denotes the element-wise multiplication between two vectors and is meant to retain only the NN-predicted stress in the solid region of the design (where the element density is 1).


\subsubsection{Vanilla DeepONet}
\label{sec:don}
\begin{figure}[h!] 
    \centering
         \includegraphics[width=0.7\textwidth]{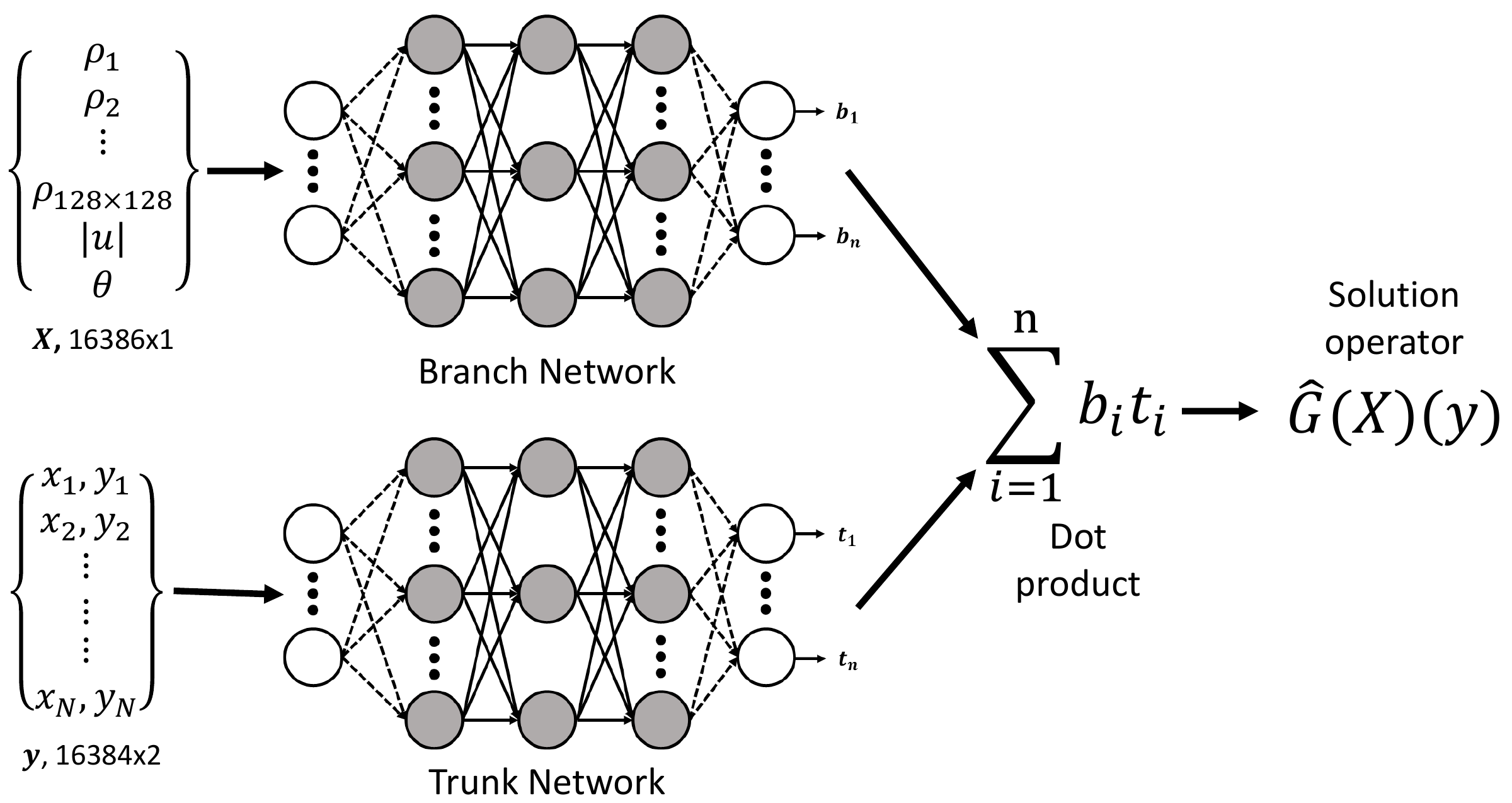}
    \caption{Schematic of the vanilla DeepONet used in this work. $N$ and $n$ denote the number of elements and the hidden dimensions, respectively.}
    \label{don_schematic}
\end{figure}

The deep operator network (DeepONet) \cite{lu2021learning} is a novel NN architecture proposed to learn nonlinear operators. It consists of a branch and a trunk NN, and a dot product is used to connect the outputs for both networks as the final DeepONet output. A schematic of the DeepONet architecture is shown in \fref{don_schematic}. We adopted the DeepONet implementation in DeepXDE \cite{lu2021deepxde} to predict the output stress field, and here "vanilla" means the unmodified version of DeepONet. The DeepONet used fully connected NNs in both branch and trunk networks. The flattened element density array (16384$\times$1) and the load parameters (2$\times$1) were concatenated together as the input for the branch network. The branch network has seven layers (including input and output) with the following number of neurons: $[16386,175,256,256,256,128,256]$. The 2D coordinates (normalized to the range $[0,1]$, 16384$\times$2) for all element centroids were inputs to the trunk network. The trunk network has seven layers with the following number of neurons: $[2,256,256,256,256,256,256]$. The output hidden dimension of both networks was set to 256, and the dot product operation combines the intermediate outputs of the two networks to produce the final output stress field. The ReLU activation function was used since the von Mises stress is non-negative. The total number of trainable parameters in the network is 3505677. The same loss function and optimizer as mentioned in \sref{sec:resunet} were used to train the DeepONet for a total of 400000 epochs.

\subsubsection{ResUNet-based DeepONet}
\label{sec:rdon}
The vanilla DeepONet, when applied to a problem with varying geometry and parametric loads, has some disadvantages. First, the input geometries are represented by a 2D, binary element density field $\bm{\rho}$. Therefore, it is expected that it carries significant spatial information about the geometry, which is key to accurate stress predictions. However, in the vanilla DeepONet model, the 2D element density is flattened in a 1D vector, then enters the fully connected branch NN. The "flatten" operation essentially loses most of the spatial information that can be best captured in its 2D form via convolution. Secondly, in the original work by Lu et al. \cite{lu2021learning}, where the DeepONet architecture was first proposed, the branch network is intended to encode the input functions, and the trunk network is intended to encode the geometry and coordinates. However, since all the input geometries are contained in a 128$\times$128 2D matrix space, the grid cell coordinates are identical for all cases regardless of the input geometry. Therefore, passing the constant coordinates to the branch (see \fref{don_schematic}) will not help encode the changing input geometry. Lastly, in the vanilla DeepONet architecture, data fusion between the two inputs only happens at the dot product, and prior to that, the branch and trunk networks remain independent. Wang et al. \cite{wang2022improved} suggested that this could lead to insufficient information fusion, thus adversely affecting prediction accuracy.

\begin{figure}[h!] 
    \centering
         \includegraphics[width=\textwidth]{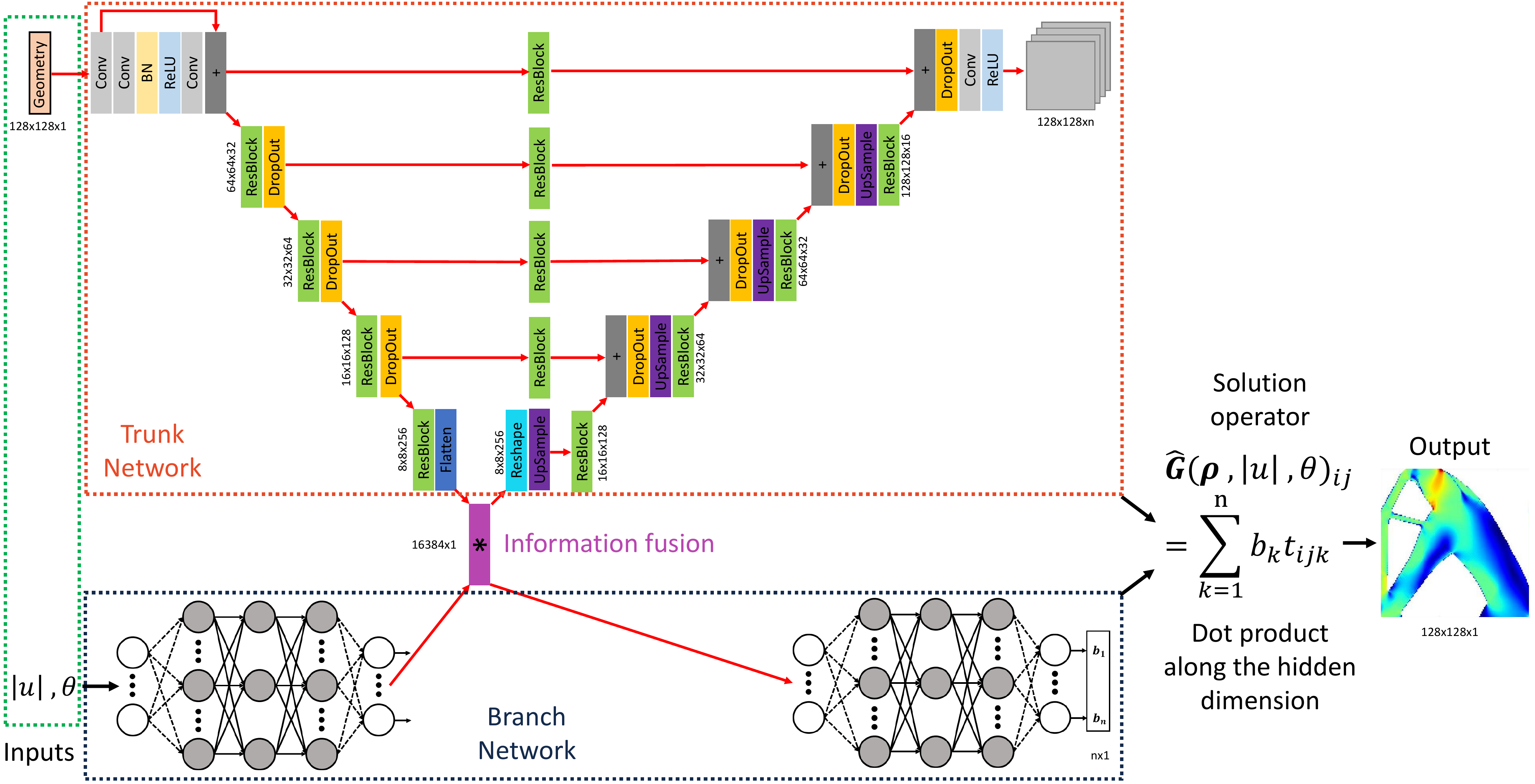}
    \caption{Schematic of the proposed ResUNet-based DeepONet used in this work. $n$ and $*$ denote the number of hidden dimensions and element-wise multiplication, respectively. }
    \label{rdon_schematic}
\end{figure}

In this work, we proposed a novel DeepONet architecture based on a ResUNet trunk network, see \fref{rdon_schematic}. It was inspired by the improved DeepONet proposed by Wang et al. \cite{wang2022improved}, where intermediate information fusion was introduced before the final dot product. The encoded load information (from the branch) and encoded geometry information (from the trunk) are mixed together via an element-wise multiplication in the latent space, and the fused information continues to propagate into the respective decoding layers in both networks. This intermediate information fusion operation is key to improving prediction accuracy. In the original DeepONet architecture, the inputs to the branch do not affect the outputs of the trunk, and vice-versa. However, mechanics knowledge implies that the input geometry and load parameters concurrently affect the output stress field. The input of one network should also affect the output of the other network to reflect this concurrent dependence. Therefore, an intermediate data fusion was introduced in the encoded latent spaces. A ResUNet was used in the trunk network to leverage convolution layers to capture and encode the rich spatial information in the input geometry, and this choice is consistent with the idea by Lu et al. \cite{lu2021learning} that the trunk network should encode the underlying geometry. To the best of the authors' knowledge, this is the first time in literature that a ResUNet is used in the trunk network of a DeepONet. A fully connected NN was used in the branch network to encode the load parameters, as the inputs are simply two scalars. Since the ResUNet trunk network generates a 128$\times$128 image with $n$ channels (where $n$ is the hidden dimension), the dot product between the branch and trunk network outputs is modified to be along the depth dimension to produce 128$\times$128$\times$1 output stress images directly, which retains the spatial information and eliminates the need to have any "flatten" operations. The proposed NN had a total of 3575056 trainable parameters with a hidden dimension $n$ of 32. Identical loss function and optimizer as mentioned in \sref{sec:resunet} were used to train the proposed DeepONet for a total of 150000 epochs.

\section{Results and discussion}
\label{sec:results}
Abaqus/Standard \citep{Abaqus2021} was used to solve all elastic-plastic problems to obtain the reference FE solutions. All FE simulations were conducted using one high-end AMD EPYC 7763 Milan CPU core. All NN training and inference were conducted using a single Nvidia A100 GPU card on Delta, an HPC cluster hosted at the National Center for Supercomputing Applications (NCSA). 

\subsection{ResUNet performance}
\label{sec:resunet_results}
\begin{figure}[h!] 
    \centering
     \subfloat[]{
         \includegraphics[trim={0cm 0cm 0cm 0cm},clip,width=0.4\textwidth]{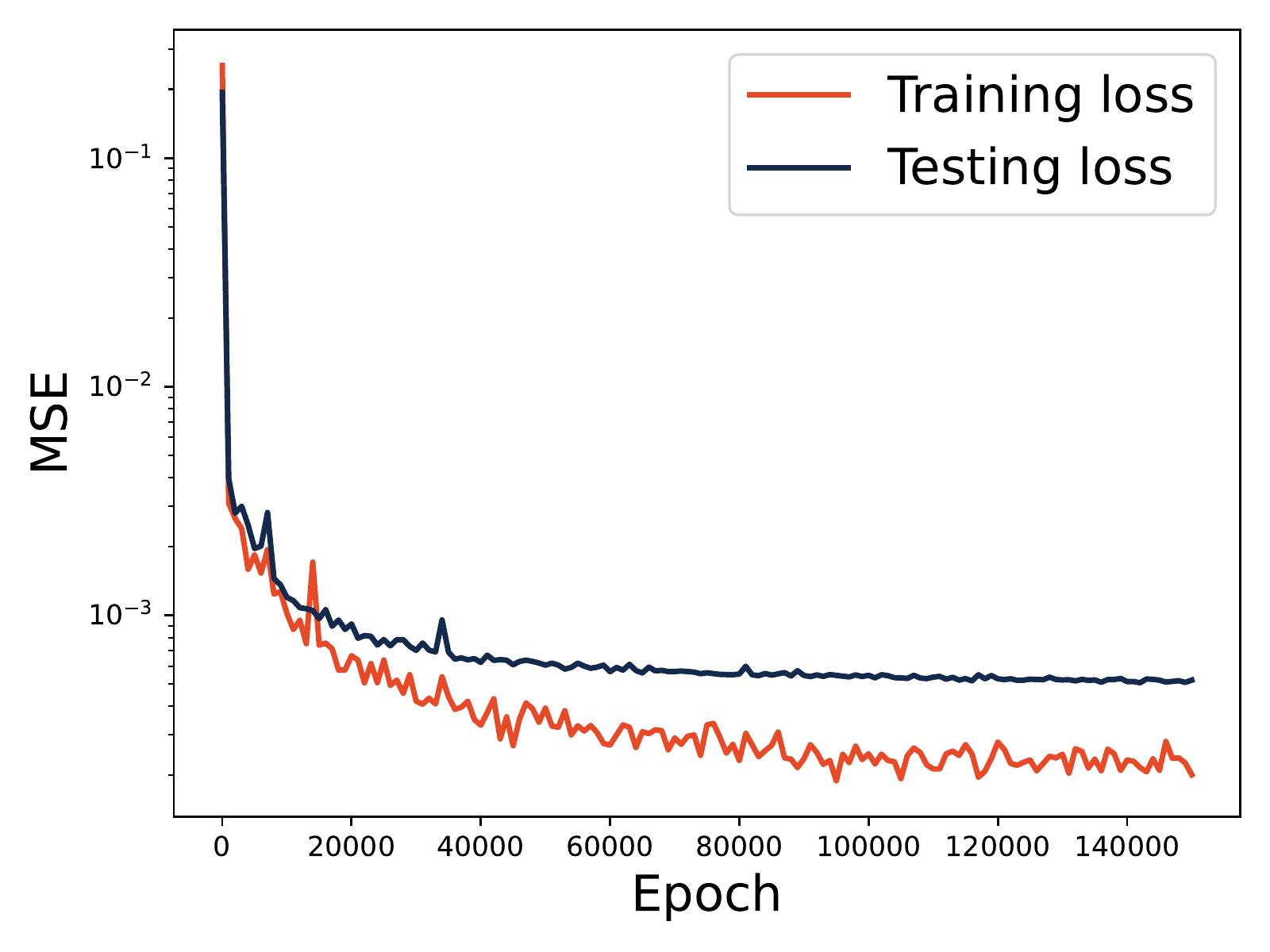}
         \label{resunet_training}
     }
     \subfloat[]{
         \includegraphics[trim={0cm 0cm 0cm 0cm},clip,width=0.4\textwidth]{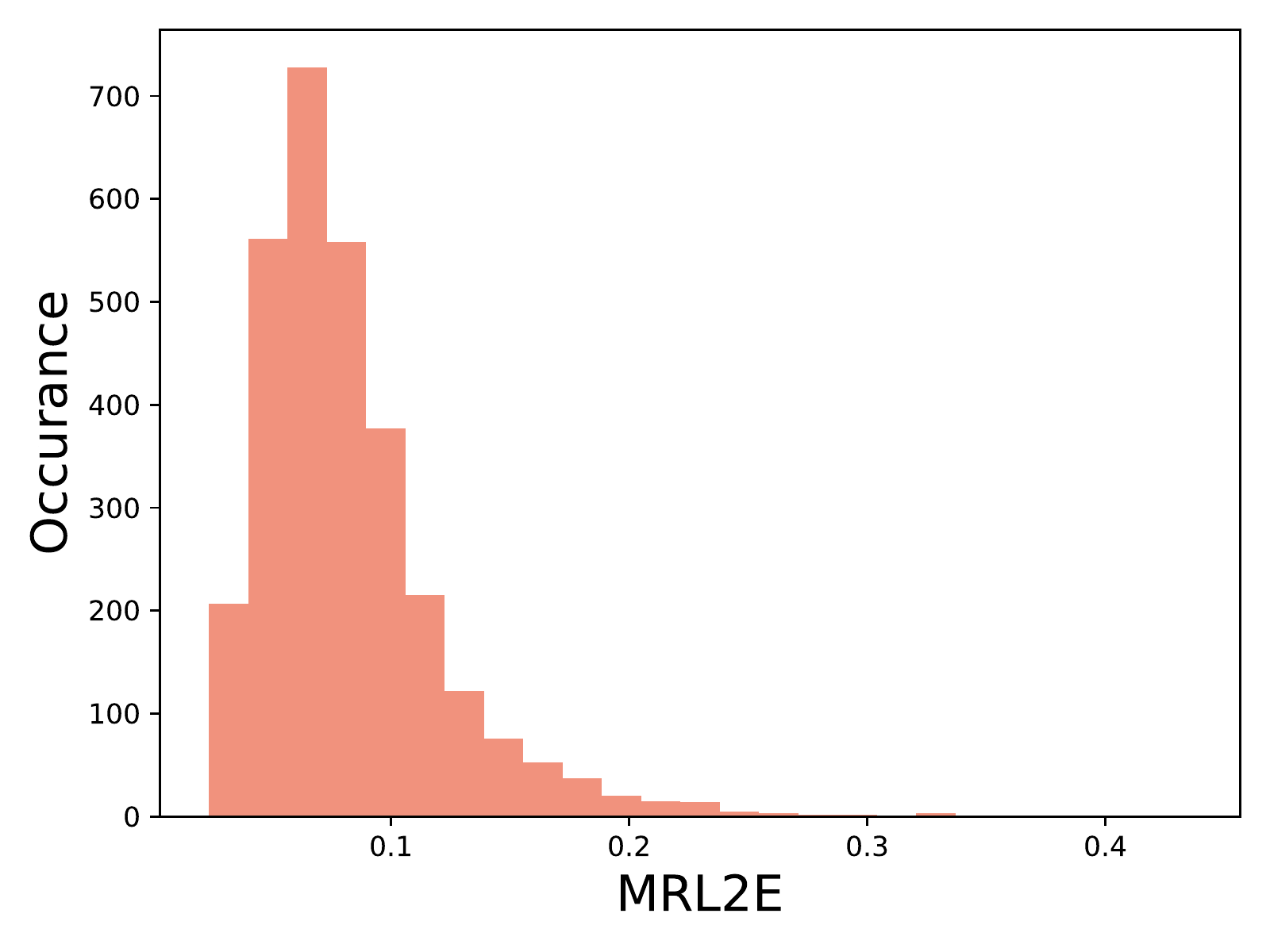}
         \label{resunet_err}
     }
    \caption{ResUNet training and error distribution: \psubref{resunet_training} Loss function value in the training and testing data sets. \psubref{resunet_err} MRL2E distribution over 3000 test cases.}
    \label{resunet_loss_err}
\end{figure}

\begin{figure}[h!] 
    \centering
     \subfloat[ Best case ]{
         \includegraphics[trim={0cm 0cm 0cm 0cm},clip,width=0.19\textwidth]{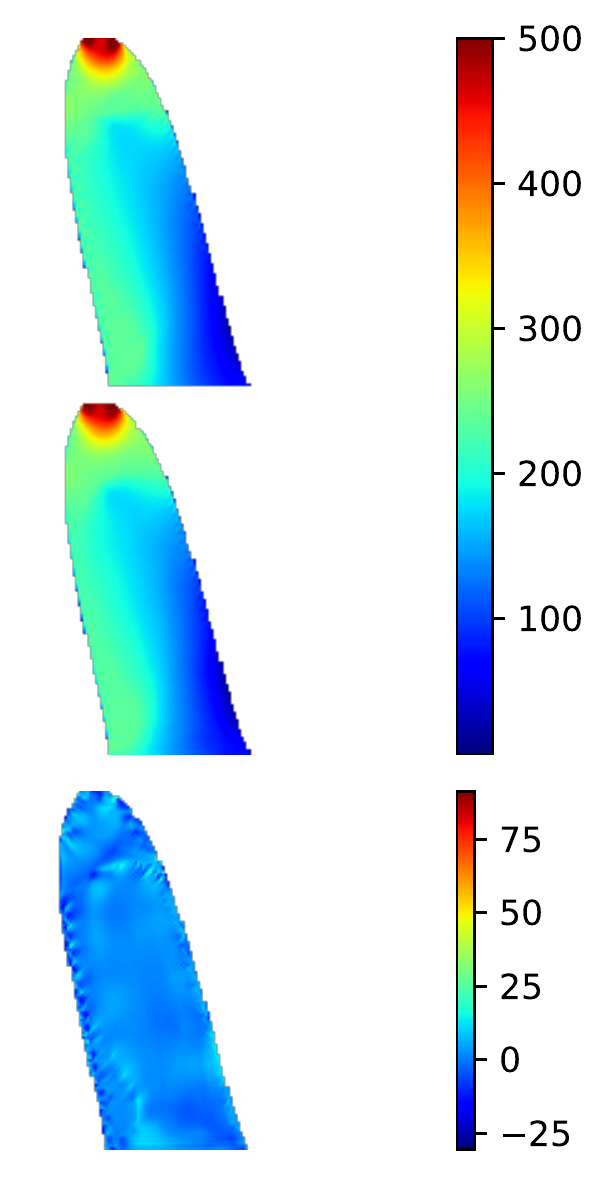}
         \label{rd0}
     }
     \subfloat[ 25$^{th}$ percentile ]{
         \includegraphics[trim={0cm 0cm 0cm 0cm},clip,width=0.19\textwidth]{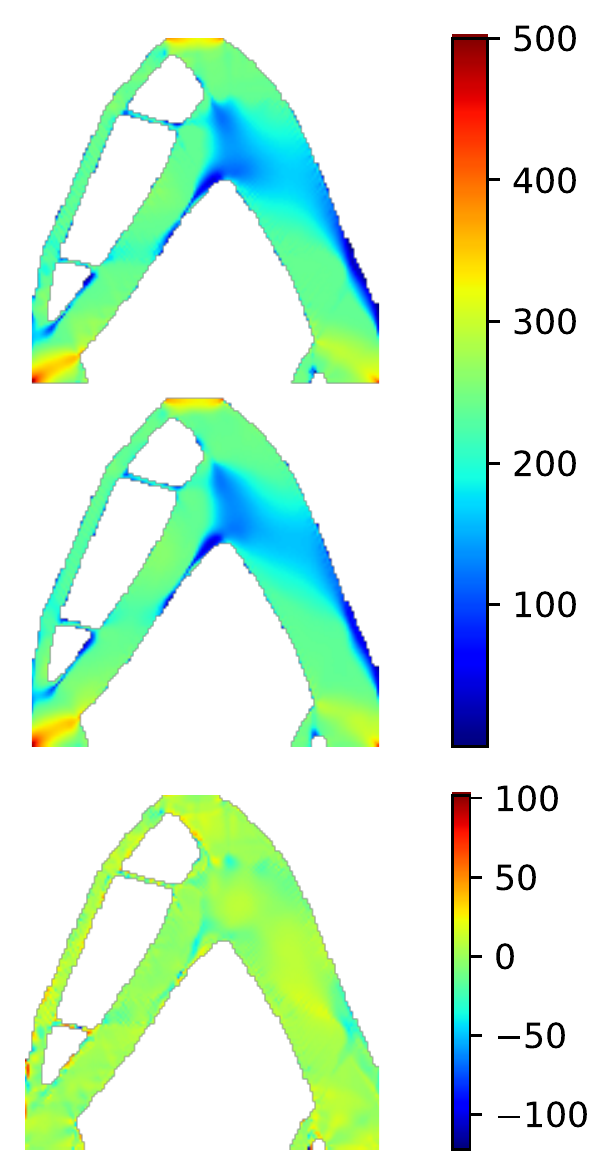}
         \label{rd1}
     }
     \subfloat[ 50$^{th}$ percentile ]{
         \includegraphics[trim={0cm 0cm 0cm 0cm},clip,width=0.19\textwidth]{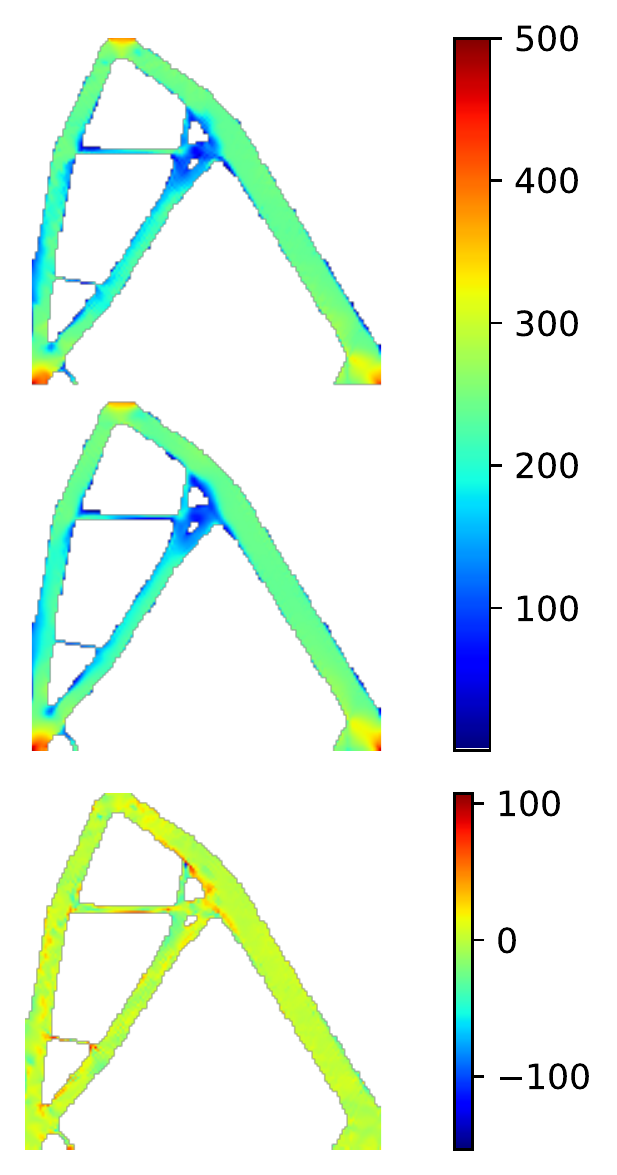}
         \label{rd2}
     }
     \subfloat[ 75$^{th}$ percentile ]{
         \includegraphics[trim={0cm 0cm 0cm 0cm},clip,width=0.19\textwidth]{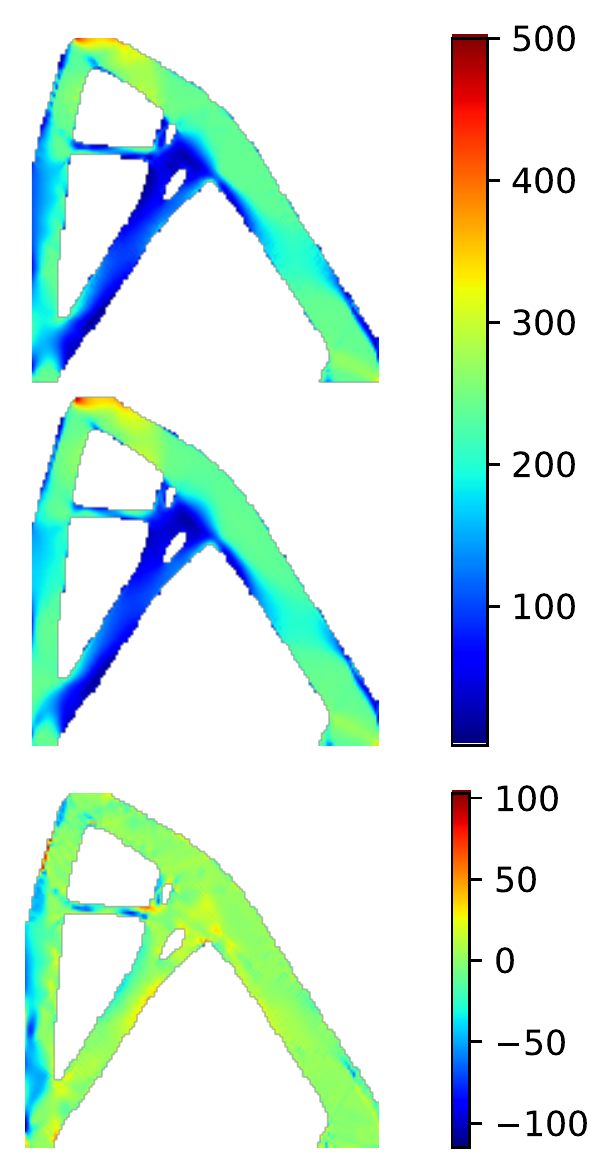}
         \label{rd3}
     }
     \subfloat[ Worst case ]{
         \includegraphics[trim={0cm 0cm 0cm 0cm},clip,width=0.19\textwidth]{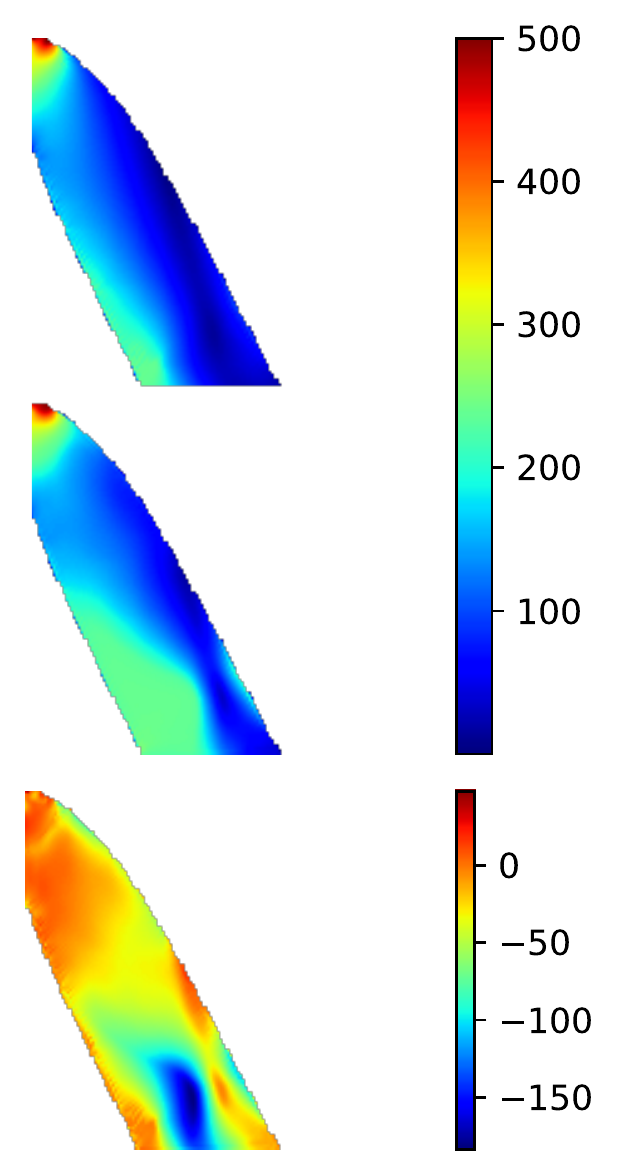}
         \label{rd4}
     }
    \caption{Selected cases of ResUNet stress predictions, ranked by percentile of MRL2E. The three rows correspond to the FE-simulated ground truth, NN predictions, and the signed error distribution.}
    \label{resunet_acc}
\end{figure}

Training of the ResUNet took a total of 3727.8s, which equates to 0.0249s per epoch. Predicting 3000 test cases took a total of 2.71s, which is 9.02$\times 10^{-4}$s per case, much faster than the FE simulation with an average of 43.4s per simulation. The evolution of the loss function value during training is shown in \fref{resunet_training}. A histogram of the MRL2E per case is shown in \fref{resunet_err} illustrating the distribution of NN prediction error within all the test cases. To show the spatial distribution of the NN prediction error, selected cases are presented in \fref{resunet_acc}. Key statistics of the MRL2E in the testing data set are shown in \tref{resunet_results}.

\begin{table}[h!]
    \caption{MRL2E statistics of the proposed ResUNet model}
    \small
    \centering
    \begin{tabular}{ccccccccc}
      & \vline & Mean & Std. deviation & Minimum & Maximum & Median \\
    \hline
    Value & \vline  & 0.0818 & 0.0390 & 0.0236 & 0.4359 & 0.0732\\
    \end{tabular}
    \label{resunet_results}
\end{table}

The ResUNet provided fairly rapid convergence during training; the performance of the model stabilized after around 50000 epochs. Judging from the MRL2E distribution in \fref{resunet_err} and \tref{resunet_results}, we see that most of the test cases have lower than 10\% MRL2E, although in the worst case, the error can be as high as 43.6\%. The contour plots in \fref{resunet_acc} show that the NN-predicted stress fields closely resemble the FE-predicted ones for cases up to 75$^{th}$ percentile MRL2E (second column from right). Even in the worst-case scenario with 43.6\% MRL2E, the ResUNet correctly predicted the location of the highest stress (at the upper left corner), which is the most critical point of the overall design and is the one that deserves the most attention during the design evaluation process.

\subsection{Vanilla DeepONet performance}
\label{sec:don_results}
\begin{figure}[h!] 
    \centering
     \subfloat[]{
         \includegraphics[trim={0cm 0cm 0cm 0cm},clip,width=0.4\textwidth]{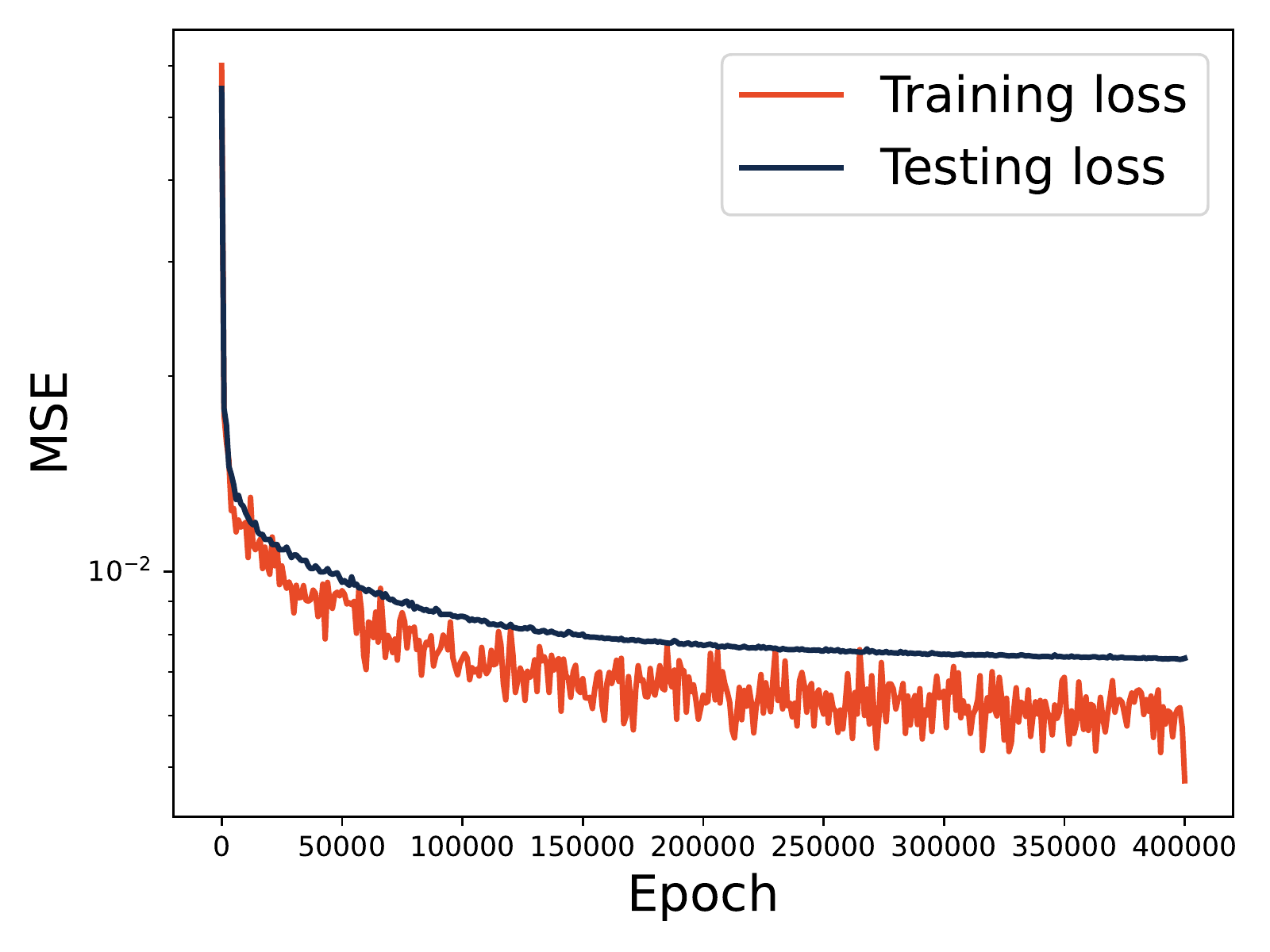}
         \label{vdon_training}
     }
     \subfloat[]{
         \includegraphics[trim={0cm 0cm 0cm 0cm},clip,width=0.4\textwidth]{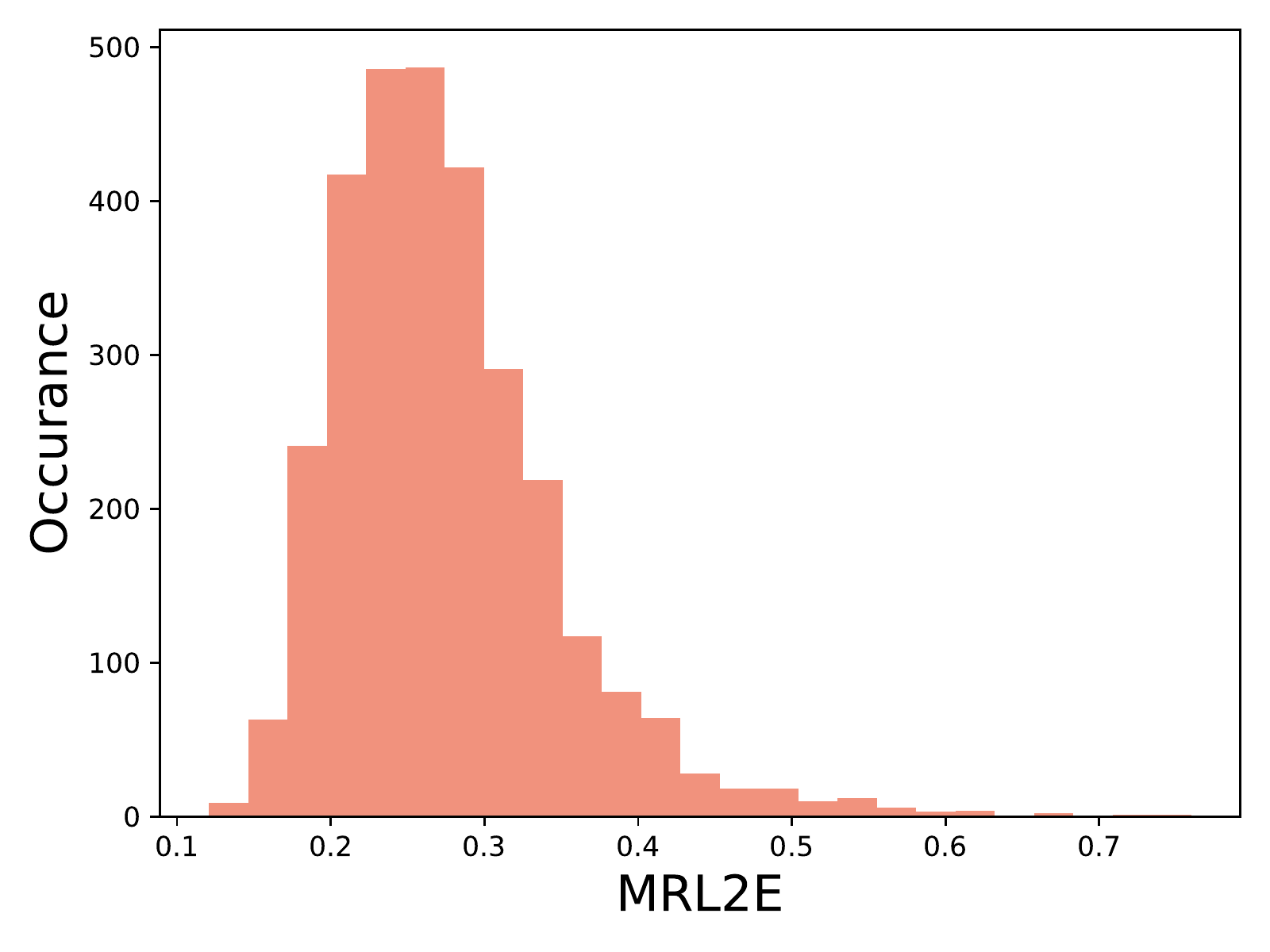}
         \label{vdon_err}
     }
    \caption{Vanilla DeepONet training and error distribution: \psubref{vdon_training} Loss function value in the training and testing data sets. \psubref{vdon_err} MRL2E distribution over 3000 test cases.}
    \label{vdon_loss_err}
\end{figure}

\begin{figure}[h!] 
    \centering
     \subfloat[ Best case ]{
         \includegraphics[trim={0cm 0cm 0cm 0cm},clip,width=0.19\textwidth]{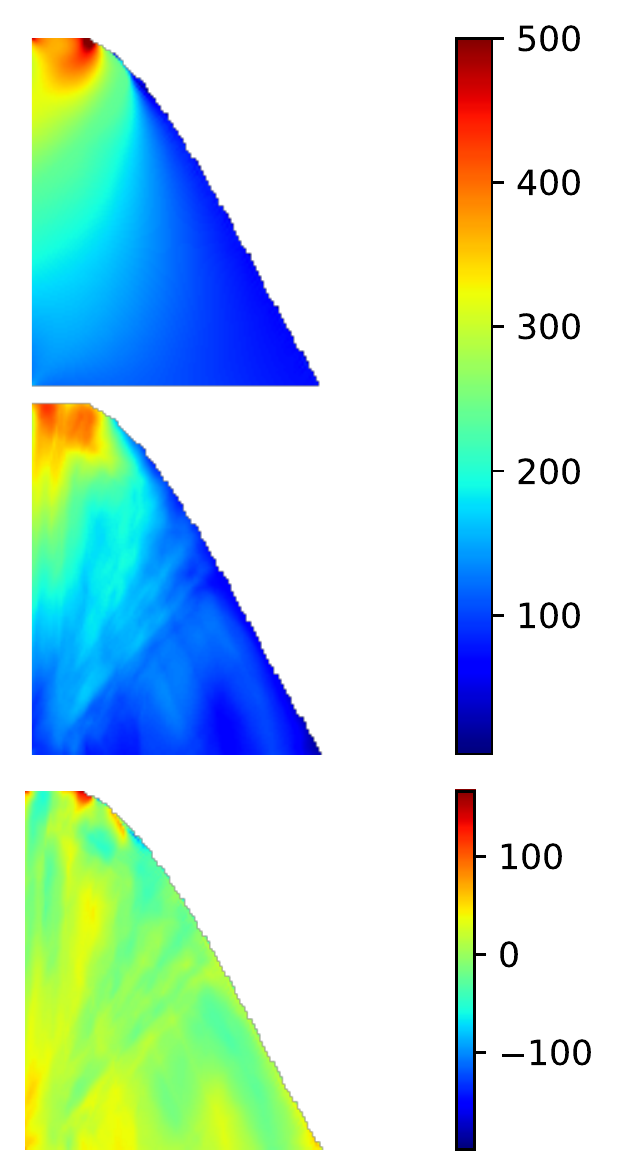}
         \label{vd0}
     }
     \subfloat[ 25$^{th}$ percentile ]{
         \includegraphics[trim={0cm 0cm 0cm 0cm},clip,width=0.19\textwidth]{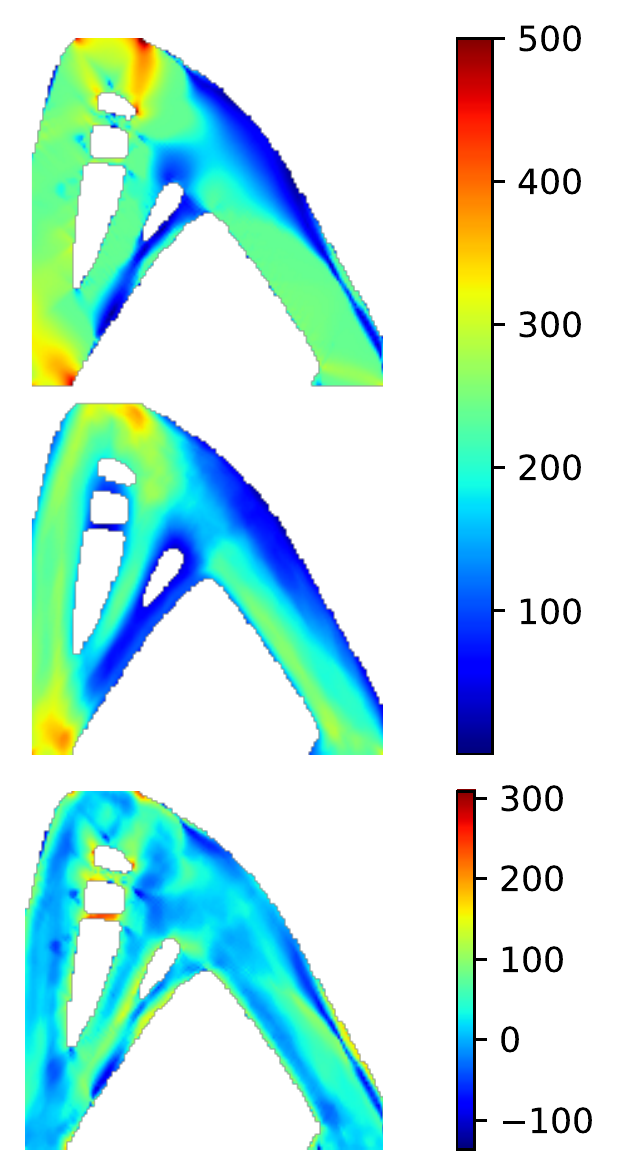}
         \label{vd1}
     }
     \subfloat[ 50$^{th}$ percentile ]{
         \includegraphics[trim={0cm 0cm 0cm 0cm},clip,width=0.19\textwidth]{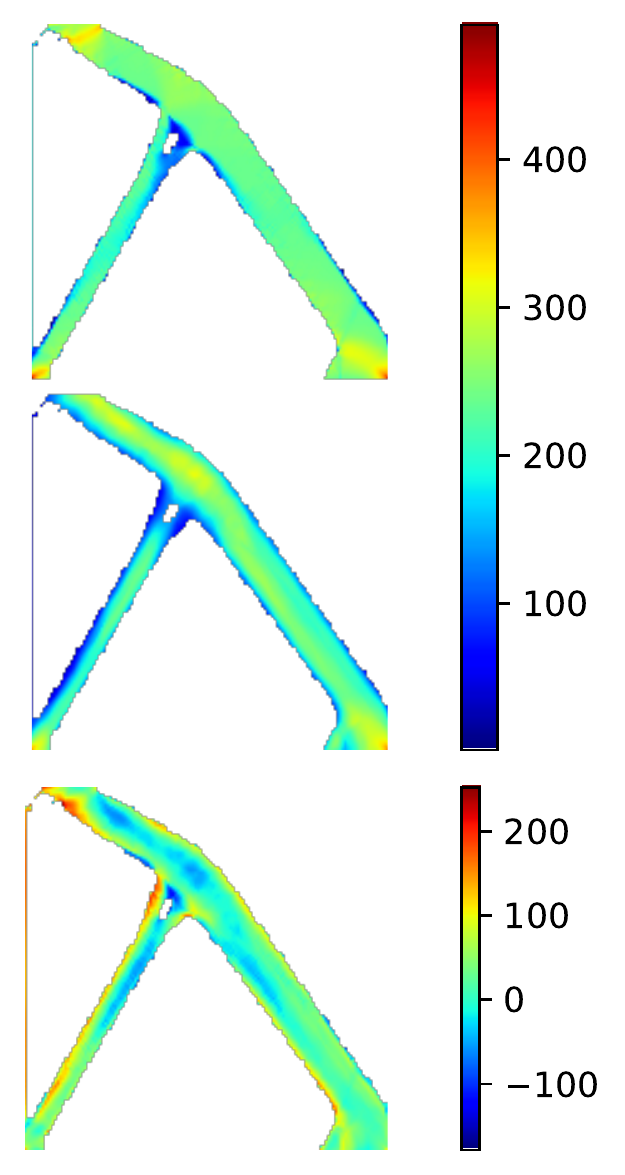}
         \label{vd2}
     }
     \subfloat[ 75$^{th}$ percentile ]{
         \includegraphics[trim={0cm 0cm 0cm 0cm},clip,width=0.19\textwidth]{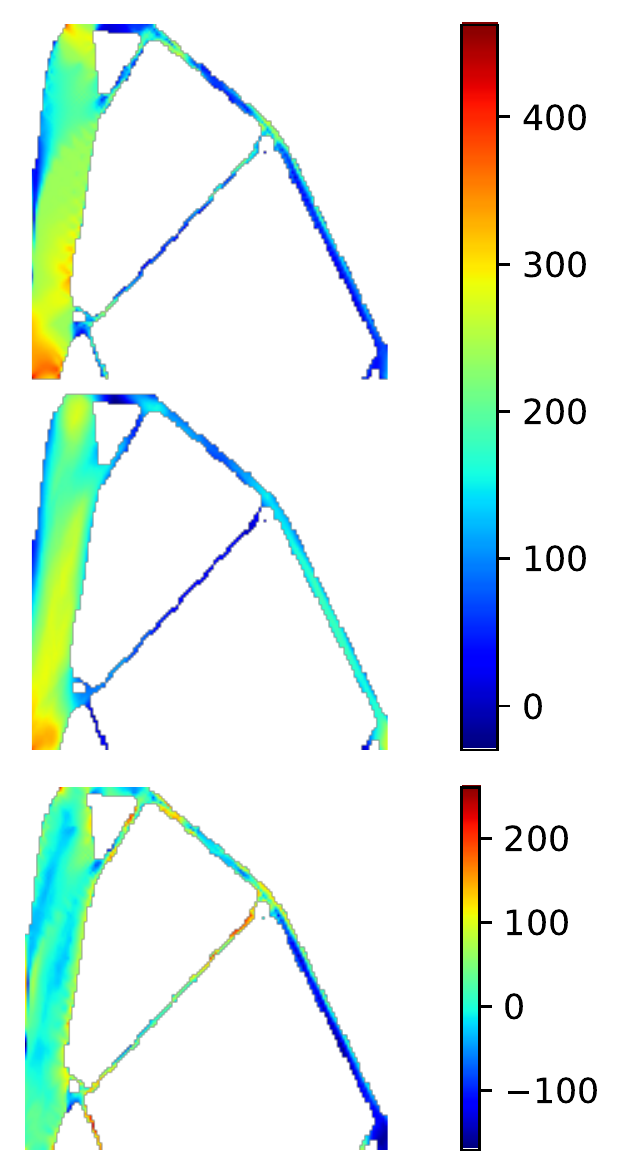}
         \label{vd3}
     }
     \subfloat[ Worst case ]{
         \includegraphics[trim={0cm 0cm 0cm 0cm},clip,width=0.19\textwidth]{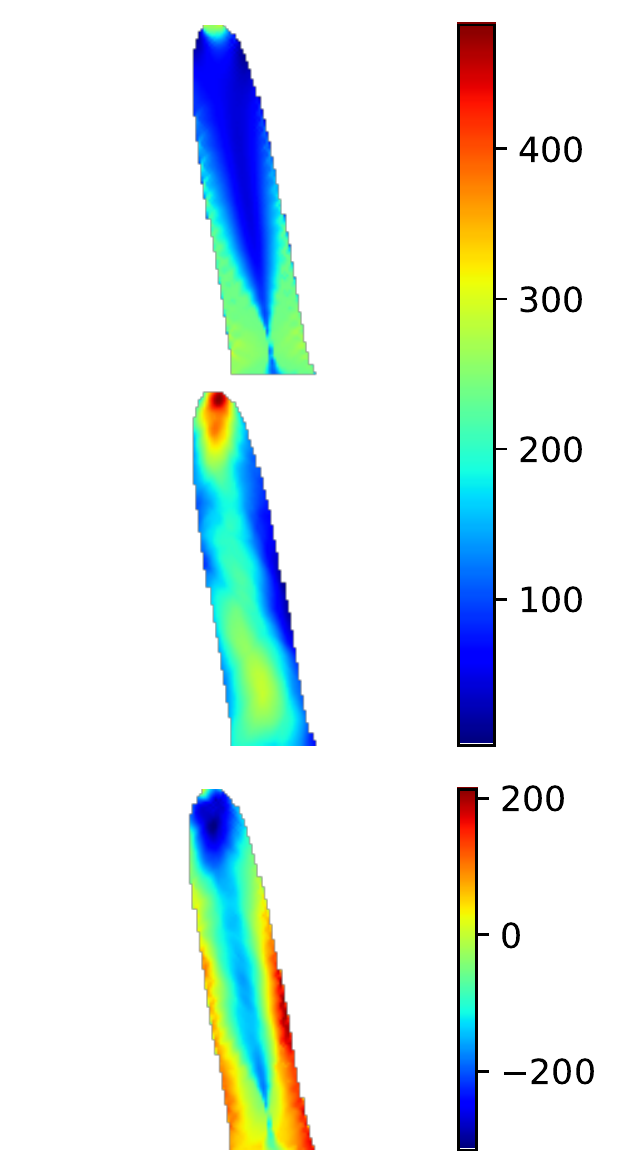}
         \label{vd4}
     }
    \caption{Selected cases of vanilla DeepONet stress predictions, ranked by percentile of MRL2E. The three rows correspond to the FE-simulated ground truth, NN predictions, and the signed error distribution.}
    \label{vdon_acc}
\end{figure}

Training of the vanilla DeepONet took a total of 1222.0s, which equates to 3.06$\times 10^{-3}$s per epoch, which is about seven times faster than the ResUNet training. Predicting 3000 test cases took a total of 0.095s, which is 3.18$\times 10^{-5}$s per case, about 27 times faster than the ResUNet prediction. The loss function history and error distribution are shown in \fref{vdon_loss_err}. Selected prediction cases are displayed in \fref{vdon_acc}. Key statistics of the MRL2E in the testing data set are shown in \tref{vdon_results}.

\begin{table}[h!]
    \caption{MRL2E statistics of the proposed vanilla DeepONet model}
    \small
    \centering
    \begin{tabular}{ccccccccc}
      & \vline & Mean & Std. deviation & Minimum & Maximum & Median \\
    \hline
    Value & \vline  & 0.2749 & 0.0723 & 0.1208 & 0.7598 & 0.2639\\
    \end{tabular}
    \label{vdon_results}
\end{table}

In the absence of any convolution layers, the vanilla DeepONet model, which consists only of fully connected NNs, trained much faster than the CNN-based ResUNet. However, since the input density field $\bm{\rho}$ contains mostly spatial information, the lack of any spatial (2D) convolution operation makes extracting these intricate spatial relations embedded in the design difficult. In addition, the trunk network of the vanilla DeepONet model (see \fref{don_schematic}) only received constant coordinate information despite the changing input geometry. Therefore, although the training was very efficient, the achieved accuracy of the vanilla DeepONet is far lower than that of the ResUNet, having an average MRL2E of 27.5\%. The contour plots in \fref{vdon_acc} show that even in the best-case scenario (left-most column), the stress concentration points on the top of the design were not predicted very accurately by the vanilla DeepONet. Therefore, the results show that a DeepONet architecture consisting only of fully connected NNs can not accurately capture the spatial information embedded in the input design and, therefore, cannot accurately predict the resulting stress field.

\subsection{ResUNet-based DeepONet performance}
\label{sec:rdon_results}
\begin{figure}[h!] 
    \centering
     \subfloat[]{
         \includegraphics[trim={0cm 0cm 0cm 0cm},clip,width=0.4\textwidth]{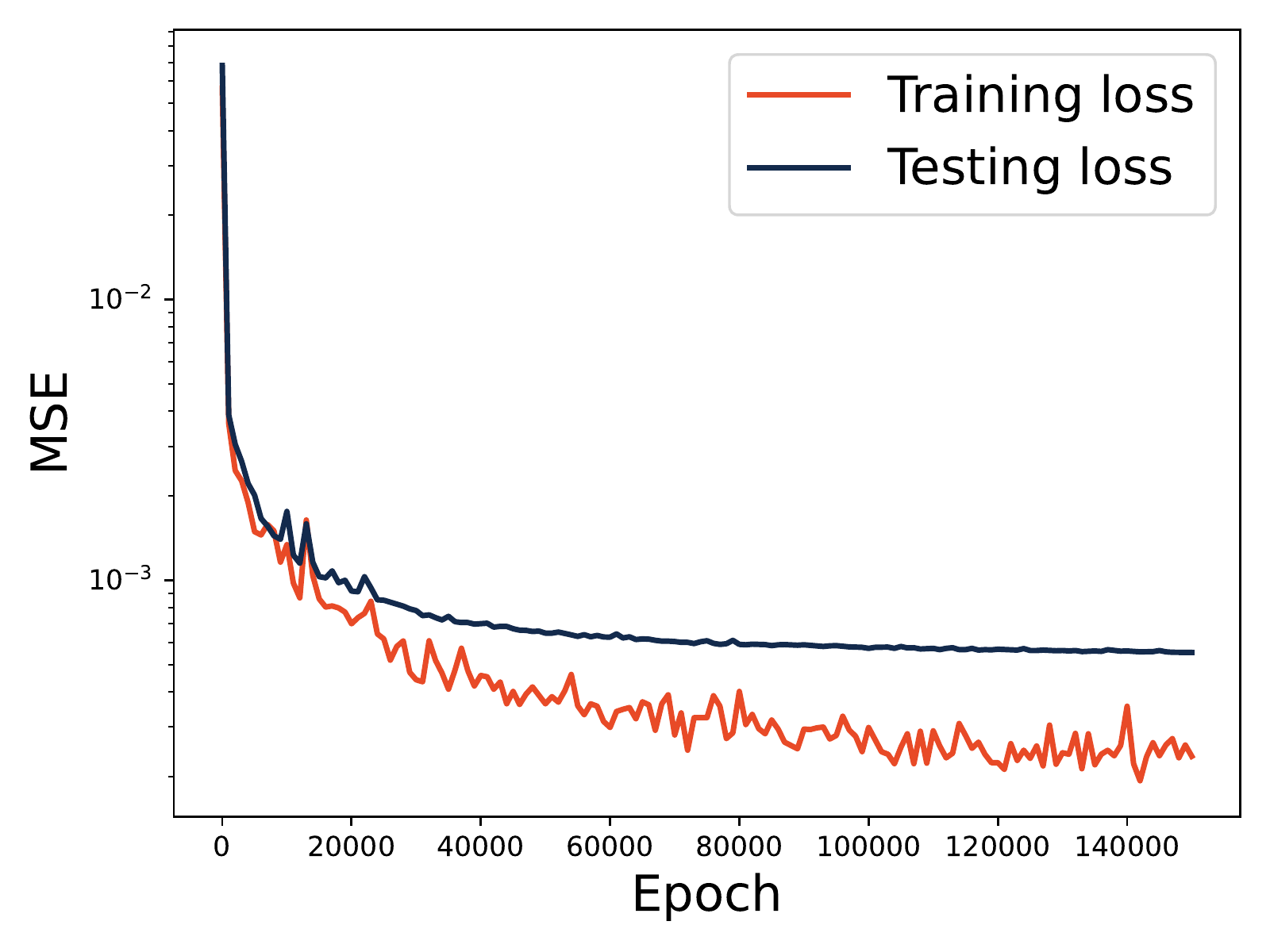}
         \label{rdon_training}
     }
     \subfloat[]{
         \includegraphics[trim={0cm 0cm 0cm 0cm},clip,width=0.4\textwidth]{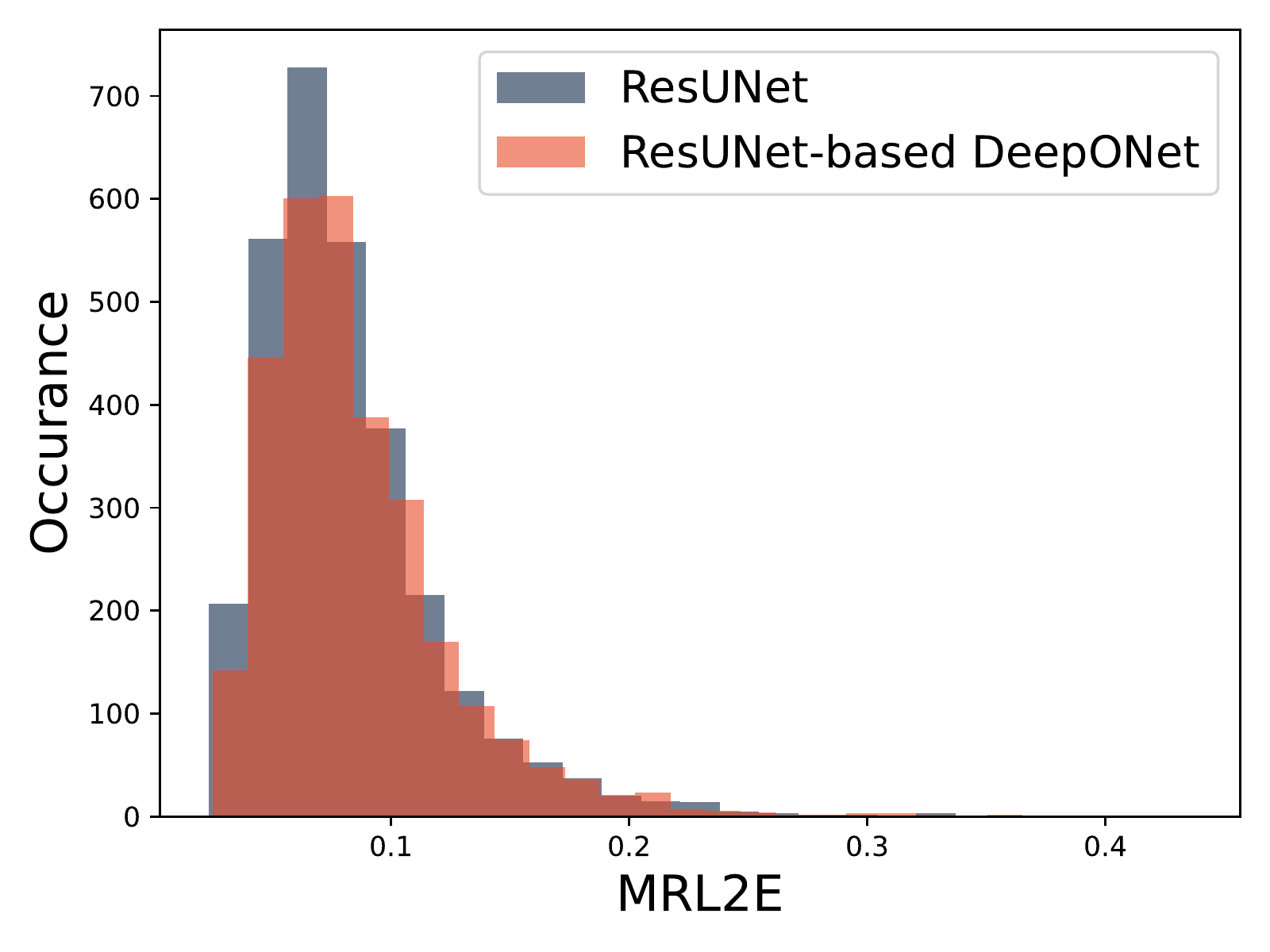}
         \label{rdon_err}
     }
    \caption{ResUNet-based DeepONet training and error distribution: \psubref{rdon_training} Loss function value in the training and testing data sets. \psubref{rdon_err} MRL2E distribution over 3000 test cases for the ResUNet and the ResUNet-based DeepONet.}
    \label{rdon_loss_err}
\end{figure}

\begin{figure}[h!] 
    \centering
     \subfloat[ Best case ]{
         \includegraphics[trim={0cm 0cm 0cm 0cm},clip,width=0.19\textwidth]{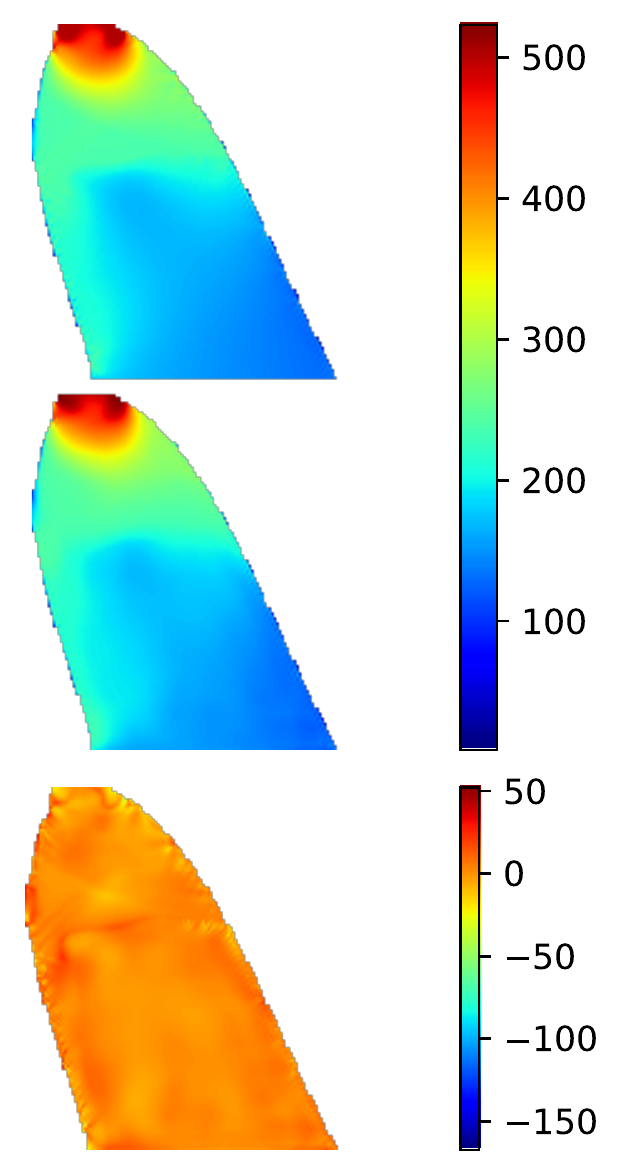}
         \label{rd0}
     }
     \subfloat[ 25$^{th}$ percentile ]{
         \includegraphics[trim={0cm 0cm 0cm 0cm},clip,width=0.19\textwidth]{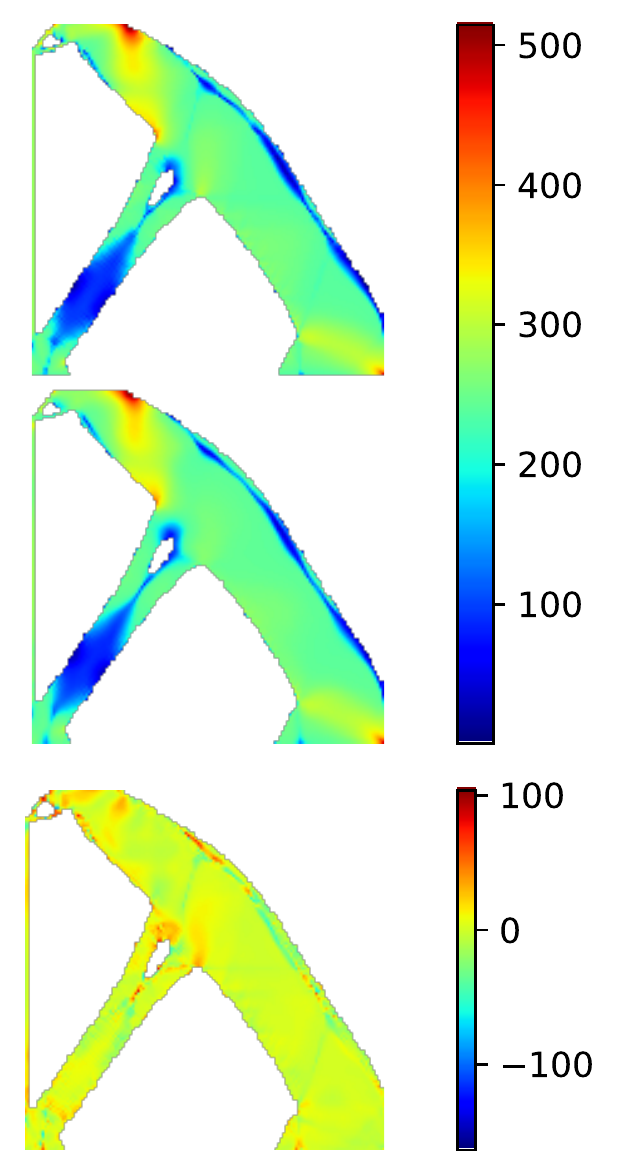}
         \label{rd1}
     }
     \subfloat[ 50$^{th}$ percentile ]{
         \includegraphics[trim={0cm 0cm 0cm 0cm},clip,width=0.19\textwidth]{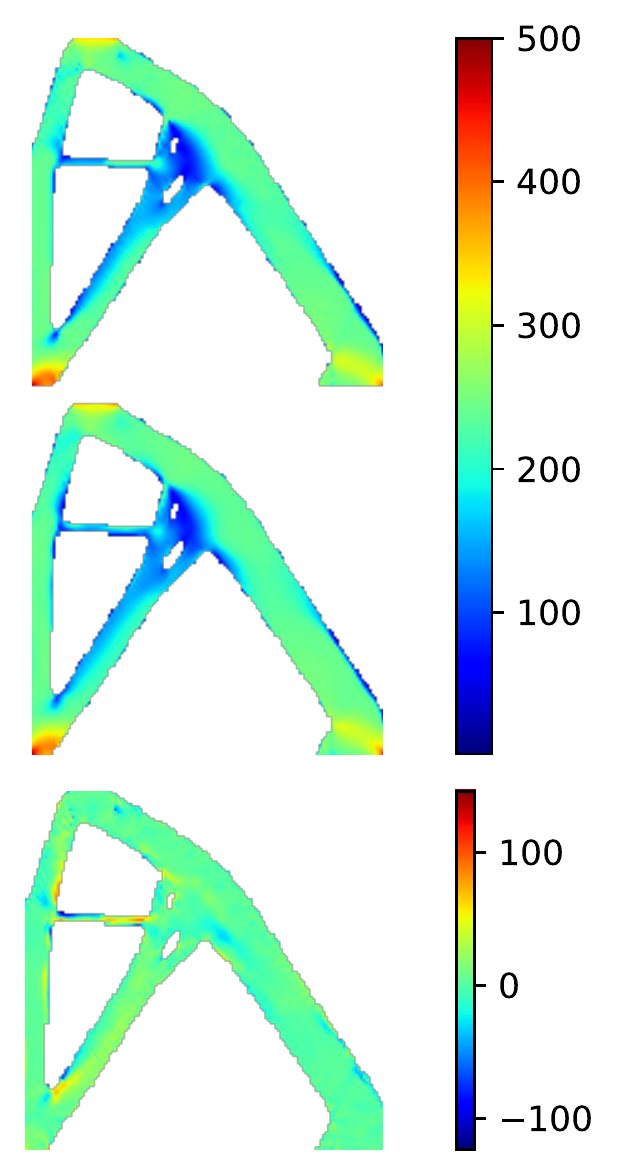}
         \label{rd2}
     }
     \subfloat[ 75$^{th}$ percentile ]{
         \includegraphics[trim={0cm 0cm 0cm 0cm},clip,width=0.19\textwidth]{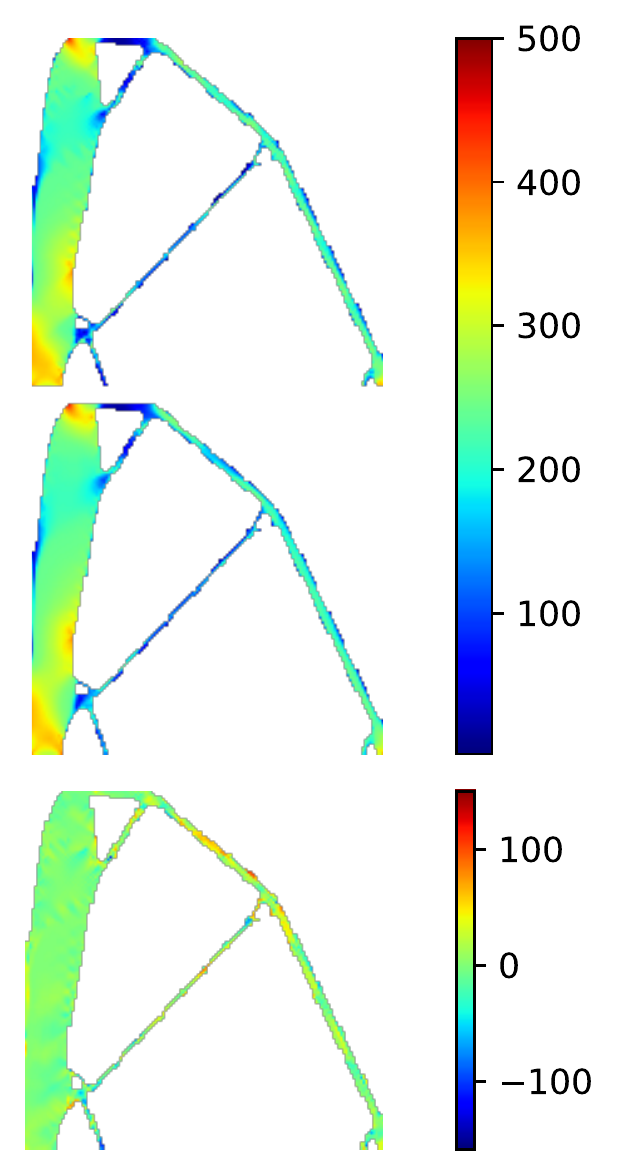}
         \label{rd3}
     }
     \subfloat[ Worst case ]{
         \includegraphics[trim={0cm 0cm 0cm 0cm},clip,width=0.19\textwidth]{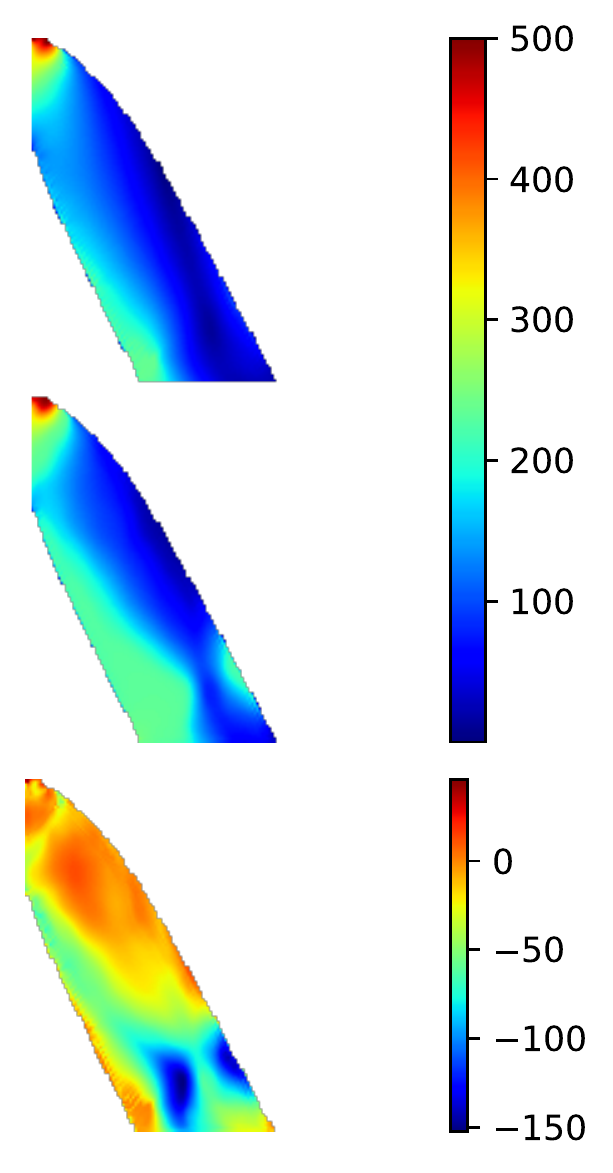}
         \label{rd4}
     }
    \caption{Selected cases of ResUNet-based DeepONet stress predictions, ranked by percentile of MRL2E. The three rows correspond to the FE-simulated ground truth, NN predictions, and the signed error distribution.}
    \label{rdon_acc}
\end{figure}

\begin{figure}[h!] 
    \centering
     \subfloat[]{
         \includegraphics[trim={0cm 0cm 0cm 0cm},clip,width=0.32\textwidth]{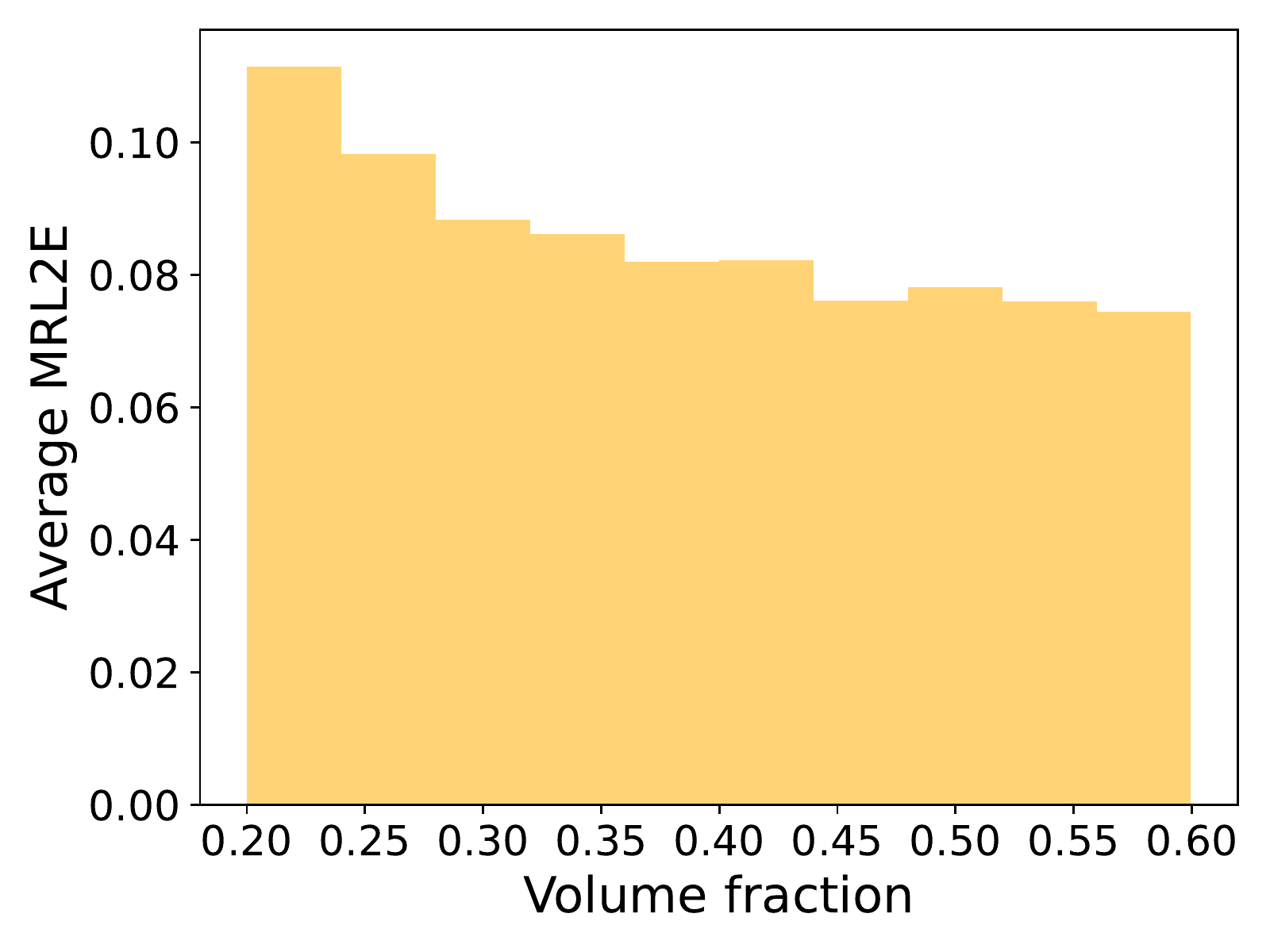}
         \label{erd0}
     }
     \subfloat[]{
         \includegraphics[trim={0cm 0cm 0cm 0cm},clip,width=0.32\textwidth]{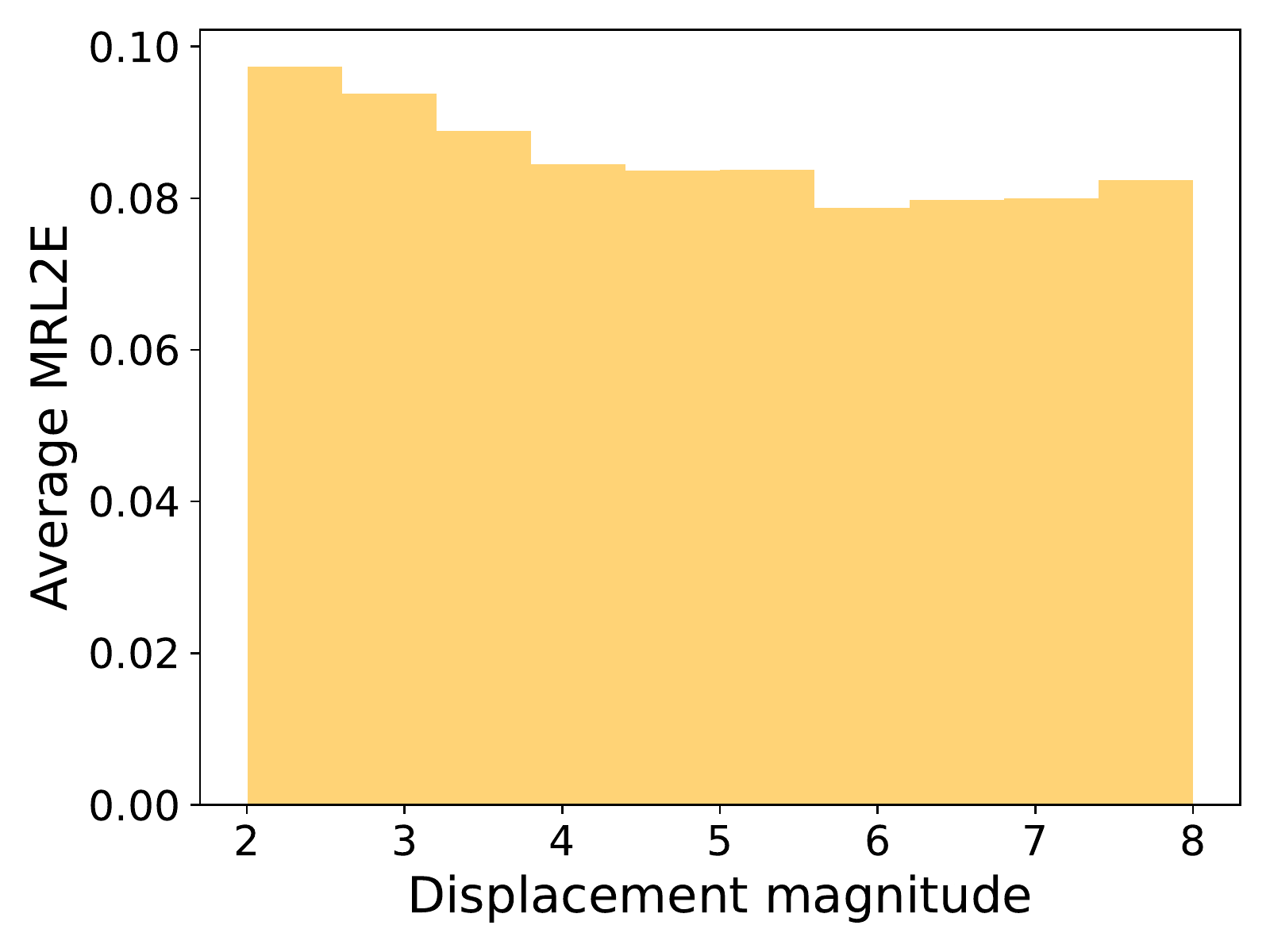}
         \label{erd1}
     }
     \subfloat[]{
         \includegraphics[trim={0cm 0cm 0cm 0cm},clip,width=0.32\textwidth]{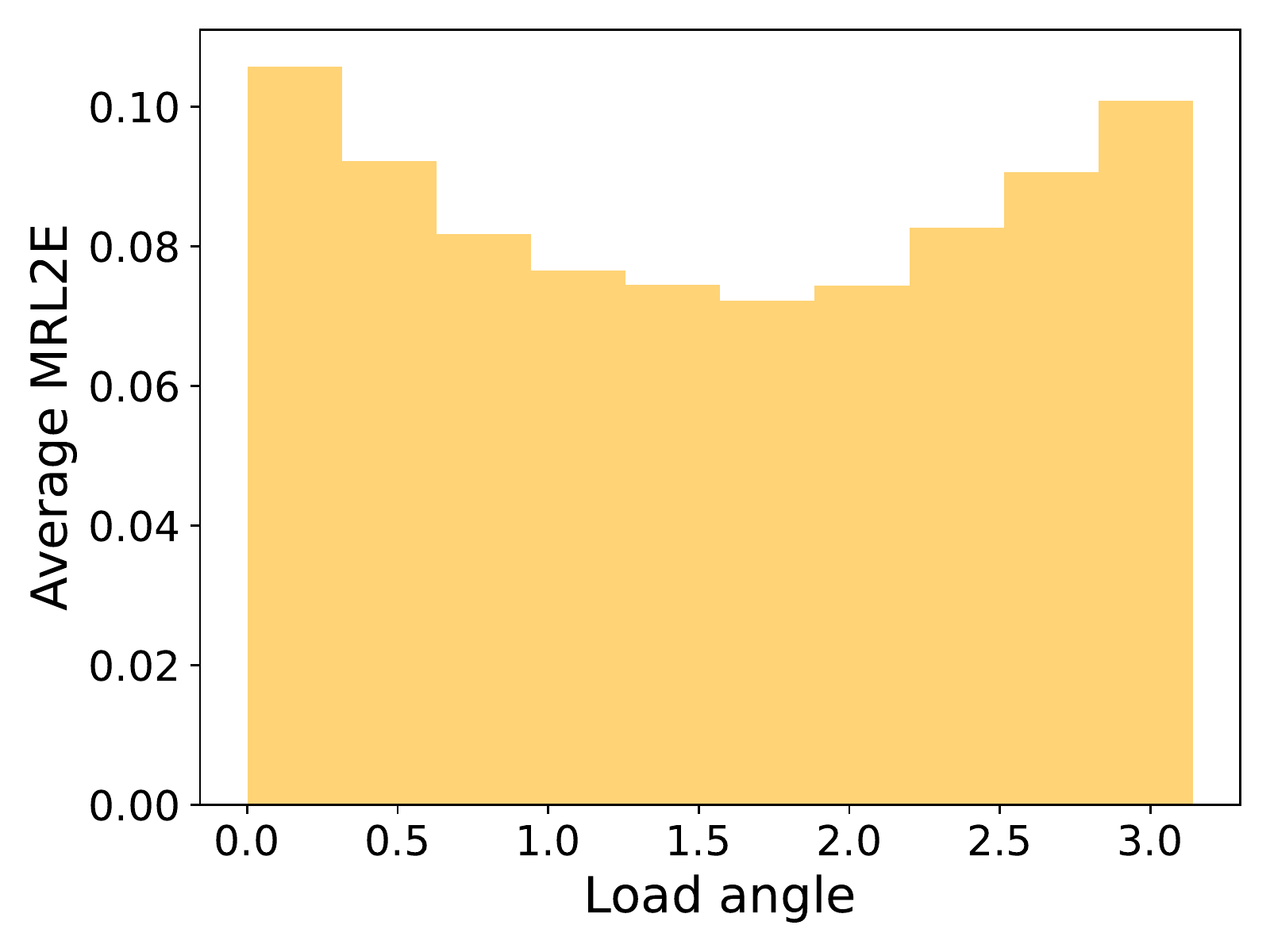}
         \label{erd2}
     }
    \caption{Bar charts showing how the average MRL2E varies with input parameters like volume fraction ($V_f$), displacement magnitude ($|\bm{u}|$) and load angle ($\theta$).}
    \label{rdon_vs_inputs}
\end{figure}

Training of the ResUNet-based DeepONet took a total of 3033.5s, which equates to 0.202s per epoch and is 23\% faster than the standalone ResUNet. Predictions took 2.37s, which is 7.91$\times 10^{-4}$s per case, five orders of magnitude faster than the FE simulations. The loss function history and error distribution are shown in \fref{rdon_loss_err}. For comparison with the standalone ResUNet model, the error distribution from \fref{resunet_err} is repeated in \fref{rdon_err}. Selected prediction cases are shown in \fref{rdon_acc}. Key statistics of the MRL2E in the testing data set are shown in \tref{rdon_results}. To show how the prediction error varies with input parameters such as $V_f$, $|\bm{u}|$ and $\theta$, bar charts for each input parameter are shown in \fref{rdon_vs_inputs}. To generate the bar charts, the input parameters (on the X-axis) were grouped into 10 different bins, and the average MRL2E over each bin was calculated and plotted.

\begin{table}[h!]
    \caption{MRL2E statistics of the proposed ResUNet-based DeepONet model}
    \small
    \centering
    \begin{tabular}{ccccccccc}
      & \vline & Mean & Std. deviation & Minimum & Maximum & Median \\
    \hline
    Value & \vline  & 0.0853 & 0.0400 & 0.0250 & 0.3947 & 0.0763\\
    \end{tabular}
    \label{rdon_results}
\end{table}

The results indicate that the NN-predicted stress field agrees well with the FE simulations up to 75$^{th}$ percentile of MRL2E, and even in the worst case, the stress concentration point is predicted correctly. The average MRL2E is 8.53\%, which is acceptable given the sheer complexity of the problem, which includes varying geometries with intricate design features and parametric loads. It is also observed that the prediction accuracy of the proposed ResUNet-based DeepONet is comparable to that of a standalone ResUNet. This observation is in agreement with a similar study by Di Leoni et al. \cite{di2021deeponet}, which also reported comparable accuracy between DeepONet and a CNN. However, the CNN in the work of Di Leoni et al. \cite{di2021deeponet} was up to 7 times faster than the DeepONet in terms of training. While in the current work, the proposed DeepONet architecture was able to train faster than the ResUNet when both have a similar number of trainable parameters. When we inspect the distribution of MRL2E with respect to different input parameters, it is obvious that the average MRL2E decreases with increasing design volume fraction. This is expected, as a larger volume fraction design typically has less intricate design features, therefore leading to fewer stress concentration points. The average MRL2E also decreases with increasing load magnitude, which is reasonable since more widespread plastic deformation at larger loads helps to smooth out local stress concentrations. Finally, the average MRL2E is symmetric with respect to the load angle, having a higher acuracy with the load is approximately parallel with the global Y-axis.

Despite the comparable but slightly worse accuracy, the proposed DeepONet architecture is advantageous over the original ResUNet as it is more memory efficient. In the original ResUNet architecture, the constant load parameters are represented as additional image channels, so each load parameter in each case is repeated 16384 (i.e., 128$\times$128) times, which is very memory intensive. While for the proposed DeepONet, the load parameters are fed into a fully connected NN without any repeating, thus saving the memory needed to hold the input data. This saving can be highly beneficial when more load parameters are needed, such as a time-dependent load path for plasticity \cite{abueidda2021deep}, which contains multiple time steps. Additionally, the DeepONet architecture allows larger flexibility over a ResUNet to handle different types of data, as it consists of two networks: the branch and the trunk. In the case of time-dependent load parameters, this can be easily tackled by the DeepONet architecture by having a recurrent NN in the branch network to encode the time dependence of the input data. The fact that the proposed ResUNet-based DeepONet can predict the complex stress field under variable geometries, parametric loading, and nonlinear material model with reasonable accuracy and high efficiency is impressive. It renders the DeepONet model a powerful candidate to rapidly predict the stress field in the preliminary design evaluation stage where high accuracy is not warranted.

\section{Conclusions and future work}
\label{sec:conc}
This work introduced a novel DeepONet architecture based on a ResUNet as the trunk network to accurately predict the full von Mises stress field given a complex input geometry, parametric loads, and a nonlinear elastic-plastic material model. The proposed network uses a ResUNet in the trunk to encode complex input geometries and extract spatial information. It includes an intermediate data fusion step through element-wise latent space multiplication to boost prediction accuracy. The performance of three different neural network architectures was compared; they are: (1) a ResUNet (baseline), (2) a DeepONet based on fully connected neural networks (baseline), and (3) the proposed DeepONet. The three networks had a similar number of trainable parameters and were trained using identical datasets and optimizers. The baseline ResUNet model achieved an average prediction accuracy of 8.2\% and can accurately predict the location of stress concentration even in the worst case. The vanilla DeepONet, although very computationally efficient, failed to achieve a reasonable prediction accuracy. Finally, the novel DeepONet proposed in this work achieved a similar prediction accuracy (8.5\%) as the ResUNet with a shorter training and inference time. The new DeepONet architecture is advantageous over ResUNet due to its memory efficiency and flexibility in terms of branch network architecture. The ability to efficiently predict stress fields in a complex geometry and material behavior under a randomly generated parametric load renders the developed deep operator network a powerful tool for preliminary design evaluation, design optimization, sensitivity analysis, uncertainty quantification, and many other nonlinear analyses that require extensive forward evaluations with variable geometries, loads, and other parameters. 

In the current work, only two load parameters were used to describe the loading. The load was assumed to linearly ramp up to the final value during the elastic-plastic simulation. In future work, more complex, time-dependent loading histories will be included via a recurrent neural network in the branch of the DeepONet. Additionally, the neural network predicts only the stress field at the end of the load step. A modified DeepONet architecture will predict the stress fields at multiple load steps in future work.

\section*{Replication of results}
The data and source code supporting this study's findings can be made available to the corresponding author upon reasonable request.

\section*{Conflict of interest}
The authors declare that they have no conflict of interest.

\section*{Acknowledgements}
The authors would like to thank the National Center for Supercomputing Applications (NCSA) at the University of Illinois, and particularly its Research Computing Directorate, the Industry Program, and the Center for Artificial Intelligence Innovation (CAII) for their support and hardware resources. This research is a part of the Delta research computing project, which is supported by the National Science Foundation (award OCI 2005572) and the State of Illinois, as well as the Illinois Computes program supported by the University of Illinois Urbana-Champaign and the University of Illinois System.

\section*{CRediT author contributions}
\textbf{Junyan He}: Conceptualization, Methodology, Software, Formal analysis, Investigation, Writing - Original Draft.
\textbf{Seid Koric}: Conceptualization, Methodology, Supervision, Resources, Writing - Original Draft, Funding Acquisition.
\textbf{Shashank Kushwaha}: Investigation, Writing - Original Draft.
\textbf{Diab Abueidda}: Supervision, Writing - Review \& Editing.
\textbf{Jaewan Park}: Investigation, Writing - Original Draft.
 \textbf{Iwona Jasiuk}: Supervision, Writing - Review \& Editing.

\bibliographystyle{unsrtnat}
\setlength{\bibsep}{0.0pt}
{\scriptsize \bibliography{References.bib} }
\end{document}